\documentclass[10pt, a4paper, twoside]{article}

\def\normalbaselineskipp{5.05mm plus.11mm}

\def\smallbaselineskipp{4.4mm}

\def\usenormalbaselineskipz{\baselineskip=\normalbaselineskipp}

\def\usesmallbaselineskipz{\baselineskip=\smallbaselineskipp}

\font\titletype=cmr10 at 15pt

\font\authortype=cmr10

\font\headingboldtype=cmb10 at 12pt

\font\normaltype=cmr10

\font\bf=cmb10

\font\it=cmti10

\font\mathitalictype=cmmi10

\font\mathsymboltype=cmsy10

\font\symbolsymbol=cmsy7

\font\symbolroman=cmr10

\font\symbolitalic=cmti9

\font\symbolsmallitalic=cmti7

\def\vex#1#2{\dimen114=#1 \dimen115=#2 \advance\dimen114 by 3.85mm \advance\dimen115 by 2.05mm \vrule width 0pt height \dimen114 depth \dimen115}

\def\rulez{\noindent\hrule width\hsize height.2mm}

\def\vskiplinez{\vskip 4.1mm}

\def\tocsectionitemz#1#2{\noindent\rlap{\bf #1}\kern 5mm{\it #2}}

\def\tocsubsectionitemz#1#2{\noindent\kern 5.6mm\rlap{\bf #1}\kern 8.0mm{\it #2}}

\def\tocsubsubsectionitemz#1#2{\noindent\kern 14.0mm \rlap{\bf #1}\kern 10.5mm{\it #2}}

\def\Ql{\kern-.1mm}

\def\Tl{\kern-.3mm}

\def\Ir{\kern.2mm}

\def\Vl{\kern-.2mm}

\def\numberedsectionz#1#2{\vskip 13mm \centerline{\headingboldtype #1\kern 3.5mm #2} \nobreak \vskip 3.3mm}

\def\numberedsubsectionz#1#2{\vskip 13mm \centerline{\headingboldtype #1\kern 4mm #2} \nobreak \vskip 3.3mm}

\def\numberedsubsubsectionz#1#2{\vskip 13mm \centerline{\headingboldtype #1\kern 4.5mm #2} \nobreak \vskip 3.3mm}

\def\addendumz{\vskip 13mm \centerline{\headingboldtype Addendum} \nobreak \vskip 3.3mm}

\def\acknowledgmentsz{\vskip 13mm \centerline{\headingboldtype Acknowledgments} \nobreak \vskip 3.3mm}

\def \referencesz{\vskip 13mm \centerline{\headingboldtype
References} \nobreak \vskip 2.2mm \usesmallbaselineskipz}

\def\referenceitemz#1{\hangindent \dimen100 \par \noindent #1}

\def\equationz#1#2{\noindent\rlap{#1}\kern11mm#2}

\def\equationvspacez{\vskip \baselineskip}

\def\twixtequationsvspacez{\vskip .5\baselineskip}

\def\twixtequationpartsvspacez{\vskip .25\baselineskip}

\def\backabitz{\kern -.1em}

\def\foreabitz{\kern .2em}

\setbox50=\hbox{\normaltype d\kern .02em)}
\def\letterz#1{\kern \wd50 \llap{#1\kern .02em )}}

\setbox51=\hbox{\normaltype d$^\prime$\kern .05em)}
\def\primedletterz#1{\kern \wd51 \llap{#1\kern .05em )}}

\def\equationforallsepz{\hskip .8em}

\def\ctimescz{\raise .37ex \hbox{\symbolsymbol\kern .3em \char"6A \kern -.1em \char"02\kern-.1em \char"6A\kern .3em}}

\def\odivbycz{\raise -.1ex \hbox{\kern .2em {\symbolsmallitalic /}\kern -.08em {\symbolsymbol \char"6A}\kern .27em}}

\def\cdivbyoz{\raise .15ex \hbox{\kern .41em \vrule depth .58ex width .04em height 1.55ex \kern-.02em \vrule depth -1.52ex width .51em height 1.6ex \kern -.54em{\symbolitalic /}\kern .31em}}

\def\varfrombrackz{\kern .1em}

\def\leftbracketz{[\kern .05em}

\def\leftbrackethspacez{[\kern .2em}

\def\rightbracketz{\kern .05em ]}

\def\hspacerightbracketz{\kern .15em ]}

\def\leftbornz{\hbox{$\hskip .2em\vrule depth .6ex width .04em height 1.6ex \kern -.04em \vrule depth .6ex width .5em height -.52ex \kern -.5em
\vrule depth -1.62ex width .5em height 1.7ex \kern -.28em {\symbolroman (}$}}

\def\leftbornhspacez{\hbox{$\hskip .2em\vrule depth .6ex width .04em height 1.6ex \kern -.04em \vrule depth .6ex width .5em height -.52ex \kern -.5em
\vrule depth -1.62ex width .5em height 1.7ex \kern -.28em
{\symbolroman (}\kern .1em$}}

\def\rightbornz{\hbox{${\symbolroman )} \kern -.3em \vrule depth .59ex width .5em height -.5ex \kern -.5em \vrule depth -1.62ex width .5em height 1.7ex \kern -.04em \vrule depth .583ex width .04em height 1.6ex\hskip .2em$}}

\def\hspacerightbornz{\hbox{$\kern .1em{\symbolroman )} \kern -.3em \vrule depth .59ex width .5em height -.5ex \kern -.5em \vrule depth -1.62ex width .5em height 1.7ex \kern -.04em \vrule depth .583ex width .04em height 1.6ex\hskip .2em$}}

\def\bp{\kern .025em}

\def\Sb{\kern-.01em}

\def\bornsepz{\kern .14em}

\def\borneqsepz{\kern .14em}

\def\doubleunderlinez#1{\setbox101=\hbox{$\underline{#1}$}
\dimen111=\dp101  \advance\dimen111 by -.1ex
\setbox102=\hbox{$\underline{\phantom{
\vrule width \wd101 height \ht101 depth \dimen111}}$}
\rlap{\box101} \box102}

\def\bracketsepz{\kern .06em}

\def\bracketpagesepz{\hskip .2em}  

\def\abbrevfiguresepz{\kern .3em}

\divide\dimen107 by 19
\setbox107=\hbox to \dimen107 {\rlap{\normaltype 000}}
\def\zeroess{\copy107}

\def\gluee{0em plus .001em}

\def\addot{$\skew6\ddot A$}

\def\bddot{$\skew4\ddot B$}

\def\jddot{$\skew6\ddot J$}

\def\kddot{$\skew4\ddot K$}

\def\bhat{$\skew4\widehat B$}

\def\khat{$K$\kern-.66em\raise.63ex\hbox{$\mathchar"0362$}\kern.1em}

\def\jp{\kern-.05em}

\begin{document}

\usenormalbaselineskipz

\hrule height 0pt

\centerline{Concepts of Physics, Vol. I no. 1 (2004)}

\vskip -4mm

\vskip7mm\centerline{\titletype Quantum Thought Experiments Can Define Nature}

\vskip 1.8mm

\centerline{\authortype Donald McCartor}

\vskip 12.5mm

\rulez 

\vskip 8mm

\normaltype

\usesmallbaselineskipz

\centerline{ABSTRACT}

\vskip 1.35mm

\noindent One would not think that thought experiments could matter to nature, for they are a humble human device.  Yet quantum mechanics very naturally frames thought experiments (as distinct from precisely defining what exists).  They exemplify the informing powers of radiation.  Though based on wave functions that have time symmetry, these tableaus inevitably tell of irreversible behavior by nature.  The paper sketches how John von~Neumann\rq{}s measurement theory fits into this and retells N.~David Mermin\rq{}s baseball story.

\vskip 2.3mm

\tocsectionitemz{1}{A law both important and imperfect}

\tocsectionitemz{2}{Perfect and imperfect ideas}

\tocsectionitemz{3}{\Tl Thought experiments}

\tocsectionitemz{4}{\Tl The probabilities of physics}

\tocsectionitemz{5}{Measurement theory}

\tocsectionitemz{6}{\Tl Typicalness, radiation, and inference}

\tocsubsectionitemz{6.1}{\Ql Quantum measurement}

\tocsubsectionitemz{6.2}{\Ql Quantum photography}

\tocsubsectionitemz{6.3}{\Ir Imaginary experience}

\tocsubsectionitemz{6.4}{\Tl Time reversal}

\tocsubsectionitemz{6.5}{Defining the behavior that is typical of nature}

\tocsectionitemz{7}{\Vl Von Neumann\rq{}s theory grounded on observation of radiation}

\tocsubsectionitemz{7.1}{Basics}

\tocsubsectionitemz{7.2}{Primitive observations}

\tocsubsectionitemz{7.3}{\Tl The two division operations}

\tocsubsectionitemz{7.4}{Four equations}

\tocsubsubsectionitemz{7.4.1}{When instrumented observations do not interfere}

\tocsubsubsectionitemz{7.4.2}{When observations are subsidiary to instrumented ones}

\tocsectionitemz{8}{Conclusion}

\vskip 3.6mm

\rulez

\normaltype

\usenormalbaselineskipz

\numberedsectionz{1}{A law both important and imperfect}

\noindent How often the things we do have unintended consequences.  Yet it could be worse.  Look at your dinner plate before you.  Whether you eat the peas or potatoes first might determine which hibiscus flower a bee visits in a far swale.  A walrus may live or die on your choice of fork to eat your salad.  Such things have not always seemed impossible.

Belief in physics has made these things seem impossible; we scorn them as magic now.  When something material passes from fork to walrus, if only a beam of light, then the walrus\rq s fate can depend on your choice of fork.  Otherwise, no.

Here a philosopher brings up an interesting point.  Should you wish to eat your dinner in peace, unconcerned that you might have an unplanned, fatal effect on walruses, physics is a disconcertingly shaky thing to rely on.  For physical theories change.  But there is no need to resort to sophistications.  Common experience suffices to teach us that we may sup untroubled.

All we need to do is gather statistics.  If we watch often enough, to see which fork is chosen and when walruses die, we will find that there is no correlation between them.  And that is all we {\it can\/} know.  To go farther and suppose that our choice of fork never kills a walrus is something so hopeless to back with evidence that the idea is quite meaningless.  For in any case one can think that sometimes it may kill a walrus, and then it works out like this:  If it does happen, as often the choice saves a walrus as kills one.  Things are balancing out.

And we are not able to tell how it will go when we choose our fork, so there is no change we can make in our behavior that will help.  Hence we may dine with a serene conscience.  Neither are we able to tell after the fact, whether we had killed a walrus.  There will be no occasion for rue.  This is our experience with salad forks and walruses. 

Now a physicist speaks up.  It might be N. David Mermin; this theme is his (Mermin [1990]\bracketsepz).  Every person, he says, will want to believe that their dinner is a truly private affair.  It will not satisfy that you will never know of all the walruses you have saved and killed with your dining habits and that anyway you could do nothing about it.  The forks on the table and the walruses in a distant sea have simply nothing to do with each other.  Period!  Any other idea is madness.

Well, in fact, the forks and the walruses could have something to do with each other in this way: they could be affected by a common cause.  For example, the sun rises and sets on both.  If the walruses on one side of the earth tend to die in the morning and a person on the other side tends to choose the shorter fork in the evening, then forks and walruses will show a correlation.  But even so, if on an occasion you should happen to choose the longer rather than the shorter fork, it will have no effect at all on the walruses.  What will happen to them will go on exactly the same way whatever you do with your forks.  That is the truth of the matter.

But the philosopher is keen.  Sir, you have taken up an idea that is unempirical in its bones.  It can not be tested.  Once you have picked up a fork there is no opportunity to rerun history to see what would have happened to the walruses had you picked up the other fork.

The physicist, however, is not convinced that this idea is unempirical.  Should not the theories that our experiences suggest be considered as empirical as the raw experiences themselves?  We have rooted ourselves in the principle that what happens has physical causes, and we have fared excellently well.  We have found magma pools that cause volcanos to erupt and ribosomes that make proteins.

Our theories say that causes are localized in time and space.  If one thing caused another thing at a distance, it is because something traveled between them.  Nothing relevant goes from fork to walrus, so the one does not affect the other.  Choice of a different fork will leave the walruses{\rq} lives quite unchanged.  Our physical theories insist upon that very thing.  This is clearly true of classical physics.  Quantum physics is still a bit of an enigma.

The philosopher is skeptical.  Are we to be electrified that classical physics consistently claims this, if there is just no way to check whether it is right in its claim?

That set our physicist aback!  Yet he came up with a riposte.  The idea that the walruses{\rq} lives would have gone on the same if you had chosen a different fork is not empty of content, he said. It has consequences.  Let me show you how.

It is possible to set up experiments in which from time to time two subatomic particles fly out in opposite directions from a source.  An instrument is put in the path of each particle, and the instruments react.  In each instrument there may be, for example, a flash in one of several chambers.  Call this, intending vagueness, an observation.  Each flight of particles ends in two events, one on either side, and sometimes they are totally correlated with each other.  This is not astonishing.  The two particles came from the same place.

Everything gets a name.  An observation with an instrument of type~$A$ is made on the particle flying out to the left.  An observation of type~$J$ is made on the right.  Observation~$A$ can end in one of two kinds of event, $a_0$~or~$a_1$.  $J$~can end in $j_0$~or~$j_1$.  What is special is that the results $a_0$ of~$A$ and $j_0$~of~$J$ always happen together.  Naturally, $a_1$~and~$j_1$ go together too.

Now we choose a variation of this experiment and propose how it might turn out.  There will be {\it two\/} kinds of observation that might be made to the left, $A$~and~$B$.  They cannot both be made on the same occasion; their instruments will get in each other{\rq}s way.  We must choose between them.  To the right there are also two separately possible observations, $J$~and~$K$.

Observation results might be correlated this way:  If $A$~and~$J$ are done, then as before the events $a_0$~and~$j_0$ always happen together.  Write these paired events as~$a_0j_0$.  So only $a_0j_0$~and~$a_1j_1$ happen.   If instead $A$~and~$K$ are done, only $a_0k_0$~and~$a_1k_1$ happen.  For $B$~and~$J$\jp, we get just $b_0j_0$~and~$b_1j_1$.  But when $B$~and~$K$ work the shift, then $b_0$~and~$k_1$ go together, and $b_1$~with~$k_0$.  In other words, $b_0k_0$~and~$b_1k_1$ never happen.  Can this be?

We shall reason about this.  The experiment is made.  Observations $A$~and~$J$ come out with $a_0$~and~$j_0$.  If we had chosen on this occasion to make the observation~$B$ on the left instead of~$A$, but had still observed with~$J$ on the right, the observation on the right would have gone the same way and given the same result that it did actually give,~$j_0$.  What we choose to do on the left, to put one or the other instrument in place, cannot change what happens on the right, no more than choice of forks can change walruses\rq{} lives.  This is so, even though both forks and walruses are affected by the sun, and both left and right by the one source of particles.

Then our left-hand result would have been~$b_0$, had we chosen~$B$, for we see that $j_0$~would still have happened, and $b_0$~and~$j_0$ always go together.

By similar logic, if we had used~$K$ as our right-hand instrument, though keeping~$A$, we would have seen $k_0$~happen.  Finally, we reckon that this means that if we had chosen differently at both places, observing with $B$~and~$K$ instead of with $A$~and~$J$\jp, then $b_0$~and~$k_0$ would have appeared.

Our hypothesis, that the events $b_0$~and~$k_0$ can never happen together, has become foul.  What we hypothesized can {\it not\/} be.  Observation results will never be correlated in that particular way.

We have reasoned on the assumption that the same will happen on one side whatever we may do on the other.  So this assumption does tell us concrete things about the world.

But perhaps it is another story if the instruments do not get in each other\rq s way?  We can see easily, without recourse to our assumption, that this snarl of correlations is impossible should we observe with $A$,~$B$, $J$\jp, and~$K$ all at the same time.  Yet even when we can do all the observations at once, sometimes we may not.

To handle that case, we may have a theory that observations $A$~and~$B$ will give the same result whether or not the other is also observed, and the same with $J$~and~$K$.  At first sight this seems enough to show that the correlations are impossible even when we observe $A$,~$B$, $J$\jp, and~$K$ just in pairs.  But it is not enough.  To show them impossible we still need that the results will go the same way despite what instruments are put in place on the {\it other\/} side. 

There are even stronger restrictions to be found on what can happen.  Hypothesized correlations, to get nixed, require no faultlessness.  For example, when we observe with instruments $A$~and~$J$\jp, the paired results $a_0j_0$~or~$a_1j_1$ might happen not always, but rather only 85\% of the time.  They still happen preferentially, they are the favored pairs.

Now $a_0j_1$~or~$a_1j_0$ happen 15\% of the time.  We can make the case more definite, if we like, by letting $a_0$~and~$a_1$ happen equally often, and the same with $j_0$~and~$j_1$.

Correlations of results of the other possible pairs of observations may be reduced in the same way, left with right: $A$~with~$K$, $B$~with~$J$\jp, and $B$~with~$K$.  Then $a_0k_0$~or~$a_1k_1$ will happen about 85\% of the time; $b_0j_0$~or~$b_1j_1$ the same; $b_0k_0$~or~$b_1k_1$ only 15\% of the time.

We are going to make a {\it long\/} series of observations with the instruments $A$~and~$J$\jp.  The experiment will run a million times.  It is not impossible---a~million seconds is short of twelve days.  This gives us a sequence of a million results at~$A$, call it~\addot, and paired results at~$J$\jp, to be called~\jddot\jp.

The percentages in these actual results cannot have strayed far from what we expected to get.  That the number of the favored pairs, $a_0j_0$ and $a_1j_1$, will lie outside the range 84\% to 86\%, has about one chance in~$10^{170}$ to happen.  (1 in 100\hskip .28em%
\zeroess\hskip\gluee\zeroess\hskip\gluee%
\zeroess\hskip\gluee\zeroess\hskip\gluee%
\zeroess\hskip\gluee\zeroess\hskip\gluee%
\zeroess\hskip\gluee\zeroess\hskip\gluee%
\zeroess\hskip\gluee\zeroess\hskip\gluee%
\zeroess\hskip\gluee\zeroess\hskip\gluee%
\zeroess\hskip\gluee\zeroess\hskip\gluee%
\zeroess\hskip\gluee\zeroess\hskip\gluee%
\zeroess\hskip\gluee\zeroess\hskip\gluee%
\zeroess\hskip\gluee\zeroess\hskip\gluee%
\zeroess\hskip\gluee\zeroess\hskip\gluee%
\zeroess\hskip\gluee\zeroess\hskip\gluee%
\zeroess\hskip\gluee\zeroess\hskip\gluee%
\zeroess\hskip\gluee\zeroess\hskip\gluee%
\zeroess\hskip\gluee\zeroess\hskip\gluee%
\zeroess\hskip\gluee\zeroess\hskip\gluee%
\zeroess\hskip\gluee\zeroess\hskip\gluee%
\zeroess\hskip\gluee\zeroess\hskip\gluee%
\zeroess\hskip\gluee\zeroess\hskip\gluee%
\zeroess\hskip\gluee\zeroess\hskip\gluee%
\zeroess\hskip\gluee\zeroess\hskip\gluee%
\zeroess\hskip\gluee\zeroess\hskip\gluee%
\zeroess\hskip\gluee\zeroess\hskip\gluee%
\zeroess\hskip\gluee\zeroess\hskip\gluee%
\zeroess\hskip\gluee\zeroess\hskip\gluee%
\zeroess\hskip\gluee\zeroess\hskip\gluee%
\zeroess\hskip\gluee\zeroess\hskip\gluee%
\zeroess\hskip\gluee000) \hbox{}  We will ignore possibilities so little likely. 

But we might have chosen to use, aport, instrument~$B$ instead of~$A$.  Would the same starboard events still have happened?  Perhaps it is not as clear to a person that this would be true for an experiment of one million runs as it is clear for one run.  If it takes six months to do them, all sorts of things could happen.  Maybe an experimenter, after looking at the early results on the left, walks over to the right and absent-mindedly spills tea on the instrument there.  Does this make any difference?

If you wish, unwanted effects can be starkly forfended.  Place the observation stations a light year away from the source on either side.  The particles will travel to them at nearly the speed of light.  Then during the whole course of the experiment no influence can go all the way from one observation station to the other, even if a year\rq s emission of particles from the source is needed.

So the sequence of results~\jddot\ would still have happened.  But what would the results of~$B$ have been?  When $B$~was perfectly correlated with~$J$ we could tell exactly.  Now we can limit things only to possibilities.  What might have happened with~$B$ at the left station is any sequence of results that is consistent with the sequence of results~\jddot\ that {\it did\/} happen, and would {\it still\/} have happened, at the right station.  That is to say, where $b_0j_0$~or~$b_1j_1$ occur about 85\% of the time, and $b_0j_1$~or~$b_1j_0$ only about 15\% of the time.

Give the name~\bhat\ to this set of credible sequences of results, which consists of those that would have been possible for~$B$ given what happened at~$J$\jp.  The set~\bhat\ is not precisely defined, of course.  But as long as \bhat~does include all sequences with a not utterly insignificant probability of happening in conjunction with~\jddot, yet even so includes no sequences with percentages too far from what is to be expected, the imprecision of its definition will cut no ice.

If we had changed the right side instrument rather than the left side one, using~$K$ instead of~$J$\jp, there are certain sequences of results that $K$~might have given us.  They are those that are consistent with the sequence~\addot\ we did get on the left side.  This set of possible sequences of results will be~\khat.

What, though, if we had chosen both a left side~$B$ and a right side~$K$?  {\it Something\/} would have happened.  And what {\it would\/} have happened is something that {\it can\/} happen.

We have reckoned that the only possible sequences for the left, had we chosen $B$~and~$J$\jp, are those in~\bhat.  And the only possible sequences for the right, had we chosen $A$~and~$K$, are those in~\khat.  Now we reckon once again, this time on the grounds that however we choose an instrument on one side, the collection of those events that might possibly have happened on the other side, had we chosen a different instrument there, must remain the same.

Accordingly, any pair of sequences that might have happened, had we chosen both $B$~and~$K$, must have one sequence in~\bhat\ and one in~\khat.  These pairs of sequences, as a group, call~$\widehat {BK}$.  And let~$\widehat {QQ}$ be those pairs of sequences in~$\widehat {BK}$ that satisfy our special requirement.  For we need, of course, that about 85\% of the time the favored results $b_0k_1$~and~$b_1k_0$ appear.  Else a pair of sequences cannot happen.

It is a simple mathematical consequence that $\widehat {QQ}$~is empty.  Choose any sequence~\bddot\ from~\bhat\ and any~\kddot\ from~\khat.  We have four sequences now, \addot~and~\jddot\ that happened, along with \bddot~and~\kddot\ that we picked.  Because of the correlation of  \bddot~with~\jddot\jp, \jddot~with~\addot, and \addot~with~\kddot, at most about 45\% of the time will $b_0$~be paired with~$k_1$ or~$b_1$~with~$k_0$.  But 45\% is far short of the 85\% that \bddot~and~\kddot\ need to meet the special requirement.

Here is where the 45\% comes from.  To get $b_0k_1$~or~$b_1k_0$ at some place in the sequences, one must break the 0-with-0 and 1-with-1 pattern.  A break will come between two sequences, between \bddot~and~\jddot\jp, \jddot~and~\addot, or \addot~and~\kddot.  For when the pattern is unbroken we will have either~$b_0j_0a_0k_0$, which gives~$b_0k_0$, or~$b_1j_1a_1k_1$, which gives~$b_1k_1$.  An example with just one break, between \jddot~and~\addot, is~$b_0j_0a_1k_1$, which gives~$b_0k_1$.  There may be more than one break appearing at the same place in the sequences, as with~$b_1j_0a_1k_0$, which uses three breaks to give~$b_1k_0$, but this is wasteful.  Because the breaks happen just 15\% of the time in any one of the three pairs of sequences, we cannot have more than $3\times 15\%$ of~$b_0k_1$~and~$b_1k_0$.

$\widehat {QQ}$~is empty.  There is {\it nothing\/} that could {\it possibly\/} have happened had we chosen $B$~and~$K$.  Then our scheme with particles does not work.  Not even if favored results come with only 85\% probability.

It seems likely that reasoning from the assumption that no matter what we do some things will be completely unaffected and will happen the same way, is something we do all the time, scarcely realizing it.  If we knew more about it we might reason this way more effectively.

There is just one more thing that Mermin mentioned.  Quantum mechanics is a well borne out theory, and according to it, the very state of affairs we have been pondering can happen---by various observations of the spins of particles fleeing in pairs in the singlet state, whose results will be correlated with the stated 85\% probabilities.  We can do experiments like that, though not yet so nicely as we would desire.

Mermin is quite unsettled by this.  The idea that to change our choice of salad fork will make no difference whatever in the lives of arctic walruses, even if both might be influenced by the sun, seems so fundamental to the nature of our lives, is used by us so often and to such good effect, that one would think that this idea could not ever lead us astray.

And how does it lead astray?  You will find naught amiss when you break bread.  Rather, build sophisticated equipment and use it expertly.  A ton of results that look totally boring will then be yours.  Only careful consideration will reveal its fell oddness.  Just in this arcane, hole-and-corner way does the idea fail us.  Even so, one would think---never.

\vskip -.73mm

\numberedsectionz{2}{Perfect and imperfect ideas}

\noindent We like to think that there will be some characteristics of the universe for us to find that are so basic that they abide exactly so.  It has almost been the definition of a philosopher to be a person who looks for those things.  Physicists are often given to doing this too.  Yet I think the state of affairs described by Mermin and retold here is substantial evidence, though of course not proof, that the universe is so subtle that there are no ideas at all about it, no matter how important, that are exactly true.

This belief or its opposite, that altogether true ideas do exist, are implicit in virtually every discussion of the significance of quantum mechanics.  No doubt Schr\"odinger, for example, when he thought that wave functions describe charge distributions, thought they did so accurately.

And I take the essence of Bohr\rq s complementarity to be just this: we must make do with an assortment of imperfectly working ideas.  He was brave to think this, yet still too cautious.  He tried to find form in this (the same form as the simplectic classical physics has) by saying that the imperfect ideas occur in pairs---complementarity.  One or the other idea will be adequate to every occasion.  But Mermin\rq s example does not suggest anything like that.

Though one could think him fundamentally right, Bohr encrusted the thought that there are no ideas to find that always work.  Following generations might have dug it out again, but this is not the sort of thing physicists like to do.

Philosophers have not shed light on this point either.  Ontology should be simply the science of existence.  But philosophers have said that the important thing is to find what exists, indeed, what {\it really\/} exists.  What {\it really\/} exists is comprehensible by ideas that are bang on the mark, of course.  Thus the science of existence is made to be based on the premise that there are fleckless ideas.

The trouble here is not that some people are thinking along these lines, for these people might still be right.  It is that the underlying posit that some ideas stand exquisitely true is not brought out.  Many philosophers (and physicists too) have taken the antithesis of the tenet that there are things that {\it really\/} exist, to be the notion that the moon is not there when we do not look---rather than the view that the idea of existence does not work to perfection.

The history of positivism has also bleared this point.  Positivism placed its bets on ideas close to experience and common sense.  But the course of physics has been contrary to that.  Great daring has paid off.  Especially since atoms were seen, this has brought positivism into disrepute as a failure of nerve.  Now, that there are some perfectly true ideas at the bottom of the universe can be felt to be a bold idea, and to find them to be our charge.  In that case, to feel there may be none would be a failure of nerve.

I can not say on oath that this would not be a failure of nerve.  But ideas that are not perfectly true need not be shallow.  They can be deep.

Still, how can believing that there may be no perfectly true ideas help us?  Is this not just resignation?  No, one has more freedom of thought.

Physicists often think in terms of ideas that they might conceive to be wholly faithful.  It could be just so that every pair of bodies attract each other in proportion to their masses and inversely to the square of the distance between them.  Newton himself, to be sure, did not believe this to be the real truth.  But others became convinced that it was, and to search for the wholly faithful ideas became science.

In quantum mechanics, the many worlds concept is an idea that if true will be absolutely true.  The entire universe will be precisely a wave function.  The same is true of the decoherence concept, at least as intended by its authors.  Wave functions are to be decohering through interaction with the environment just as precisely as classical particles would be jiggled by their neighbors.

The taste for completely competent concepts has carried through from the classical age right into the quantum age.  More impacting on physics, though, than the search by some scientists for the kind of idea that might be perfectly true, is a common, thorough repugnance, among people of all kinds, for any notion that might intimate that there are no such ideas to be found.  Even physicists who feel sure that {\it we\/} will never find these spotless ideas will share the horror.

This antipathy has scrubbed the development of physics in important directions.  The remainder of this paper will explore avenues left to weeds.

\numberedsectionz{3}{Thought experiments}

\noindent From my views on perfect and imperfect ideas it may appear that here is one of those people who are fond of chiding the rigidity of those who like clarity in thinking.  This is not really so.  If mathematics had been blessed with that simple, clear form Hilbert hoped it would have, before G\"odel, I would have been extremely contented.  Indeed, when I first looked at quantum mechanics I tried to find its meaning in ideas that would be perfectly true.

The form of quantum mechanics, with its spreading, entanglement, and probabilities, suggests that wave functions represent information rather than existents, as Heisenberg soon noticed.  Taking quantum mechanics to be an absolutely fundamental theory, I tried to think how it could represent information perfectly.  But whose information?  Under what circumstances?  To put it honestly, even under the best of circumstances information is not quite solidly defined.  Always, someone may have slipped some loaded dice in on you.  The mismatch between information as we experience it and those precisely defined mathematical wave functions is genuinely terrible.

Bohr, for his part, thought quantum mechanics fit to describe certain experiments.  They breed particles that fly through space and get detected.  To Bohr, this narrow duty is all that quantum mechanics handles in proper fashion.  That, naturally, raises the question of what special significance these experiments can have to nature, what are we to learn from them, when they are the sole legitimate object of nature\rq{}s most basic laws.  Bohr never said.  However that may be, experiments cannot be picked out cleanly from the big sea of events.  But then, the wave functions appropriate to them cannot be a precise truth.

If quantum mechanics is its lamp, then nature does not have the clear, pure form that I had looked for.  So what form does it have?

Quantum mechanics is most adroit at representing detection of radiation, I~believe.  Scattering theory is not so much about collisions as about snaring broadcast particles---this is what Born invented his probabilities for.  This is what Bohr\rq{}s favored experiments are, too.  But catching radiation is only a small part of the activities in the universe.  Whatever does this mean?

The detection of radiation can define the lawfulness in nature\rq{}s behavior.

For we have found that if we wish to learn about a {\it kind\/} of thing, we can find wave functions that {\it typify\/} it.  By this artifice (a wave function will not represent an actual example) we can calculate probabilities (also artifices) for detecting sundry possible patterns of radiation.  In a nutshell, from a thought experiment we will find what that kind of thing typically looks like.

We learn from our experience.  Sight is an ample part of that experience.  From these notional scenes we can infer, by empirical reasoning, how that kind of thing typically behaves.

But the empirical reasoning the thought experiments lead to cannot be based only on the sights produced.  Architectural likenesses between the wave functions and what goes on in the world must also come into play.  This, inescapably, is the basis for choosing a wave function that is typical of something.

To understand what these thought experiments are all about, it will be meet to set out on an inquiry into how we do physics, or, better, into how we make our way in this world with our ideas.  Particularly, our most common ideas need to be brought to the surface.  We will start by looking at what existence and logic mean to us and to classical physics.  Then we will go into what perplexes us about the thought experiments.

Some centuries ago the microscope first showed us germs.  Control over many diseases ensued.  A meticulous knowledge of the structure of metals has given us fine alloys.  We learned to look for ever more detailed descriptions of the things that exist.  This serves us handsomely.

We have ideas about what exists that are not bad, and we are finding ever better ideas.  It is by our new ideas that we see why the old ones worked as they did, for in some ways they came near to the new.  Yet what happens when we push this to the limit?

There would have to be perfect ideas of what exists at the end of the road, resemblance to which, we should think, explains the success of our imperfect ideas.  We might take the exact descriptor of what exists to be classical mechanics, as an example.

What we will then need from classical mechanics is a theorem that in a universe obeying its laws creatures that believe in classical mechanics and our customary ideas (meaning, anyway, that they act as if they believe in them) will prosper.  Note that these creatures are little chunks of that universe.

But there can be no such theorem.  A complex classical mechanical system will behave in almost any way, following on the initial conditions, so long as it avoids becoming simply chaotic.

Rather than to frame the existence of the entire universe, the genius of classical mechanics must be to loosely match initial conditions to the various situations we encounter, in fact to be used.  Just as we use it.

We most naturally think of classical mechanics as being like a shallows of a stream we have forded, something that is there that we have managed to bend to our ends, but surely the reason it is there has nothing to do with that.  To the contrary, its utility is classical mechanics{\rq} most salient characteristic.  This is quite remarkable, but, in truth, it is remarkable that {\it any\/} of our ideas are the slightest use to us.

To our green eyes it seems like this:  The sky {\it is\/} blue.  That is why our idea that the sky is blue works.  But not really.  Even if the sky is blue, this does not at bottom tell why the idea of it helps us.  (Mind the {\lq}no such theorem{\rq} above.)

We may now draw these several conclusions.  Classical mechanics can do its duty only in a world fundamentally unclassical.  If you take it with a childlike total acceptance, classical mechanics will turn upon you and deny that belief in itself ought to be useful, indeed just when that belief is utterly correct.

The concept of existence, on which classical mechanics is founded, we instinctively take to be the gold standard, equivalent to truth.  Yet it will be an idea of grave moment that nevertheless, like all ideas, serves not quite thoroughly.

Truth is not the absolute guarantor of utility, though our consciences proclaim it.  And we no more fathom why our ordinary ideas work than why wave functions do, though we suppose we do.

Just as we tend to put too much weight on existence, so with logic.  That is the moral of the following tale.

Think about the classical laws of hydrodynamics and the ocean.  These laws will tell you how a fluid will evolve if you are given the motions of all its parts at some instant.  But though you may presume if you wish that the dancing ocean is obeying the laws of hydrodynamics, still you cannot directly use these laws to predict waves. There is no time when you can know the motions of the water so well that you may precisely foretell the future. But that is what true use of the hydrodynamic laws entails.

However, physicists will infer from the laws of hydrodynamics a serviceable law of breakers.  It tells how waves nearing on a gently sloping shore will break.

That does not follow from the laws of hydrodynamics by logic.  It is not like proving the Pythagorean theorem, given Euclid\rq{}s axioms.  The inference is made by inventing simple models that seem typical of seashores and playing them out using the laws of hydrodynamics.  Then you look for generalities in the results, checking the generalities against experience to make sure you have not got off the track.

If you go to the beach the law of breakers, unlike the laws of hydrodynamics, will carry you happily, for you will know to expect breakers.  But of course you might run into a tsunami instead.

The laws of hydrodynamics, then, seem to play in the imagination.  The law of breakers is inferred from the laws of hydrodynamics, but not by means that are totally logical.  The law of breakers is practical, to be brought to bear, not without fault, immediately on events.

Still, it is common to think of the practical laws as if they were theorems that are logically derived from the axioms, the fundamental laws.  Even when it is recognized that in practice they are not proved as theorems, physicists tend to think of the more special laws as nevertheless {\it potentially\/} logically derived.  If physicists get them otherwise, then it is because they take shortcuts, for physics is difficult.  Ah, but the practical laws are not derivable by logic.

Truly then, the methods by which physicists derive practical results from principal laws are deep workings.  They are not to be taken for granted, even if it be that all our practical thinking is akin.  These workings must be the essence of the meaning of the principal laws.

Now for our disconcerting thought experiments.

A typical example of a steam engine will be, in our daily lives,  an actual steam engine that is like most steam engines in the ways we consider significant.  We judge that every steam engine can be rendered soundly by a classical physical construct.  To typify steam engines with the help of classical physics, we need only choose one of its constructs that could describe an actual steam engine, one that exists and is typical.

For use by the thought experiments, there ought also to be a wave function that will typify steam engines.  Wave functions, though, do not have the right kind of structure to render existence soundly.

You know the difficulties.  Wave functions spread and entangle, so the properties of things are defined less and less sharply as time passes.  They give us probabilities that we have not been able to assign to the flow of events in properly general fashion.  Heisenberg and Bell have taught us that wave functions have much the form of an incomplete knowledge of a physical system, but without having that form exactly.

Wave functions are capable of great depth of definition, down to the quarks.  Can there really be devices so finely descriptive of what real things are like, and yet not telling what real things are?  It appears there can, however it be that typicalness and existence become unexpectedly parted.

A switch engine steams up on the siding, and we are pretty sure what kind of thing it is.  But is it the same kind of thing as that mallet chuffing on the main line?  Well, it depends.  In the same way, a particular wave function may be typical of {\it various\/} kinds of things.  Who knows which of its details will matter to what concerns?  Wave functions must cover all these subtle possibilities.  This is uncanny.

These thought experiments based on wave functions will seem eccentric.  They declare, \lq\lq{}I was made to be a thought experiment.  That is all I can be.\rq\rq{}  We cannot say, however, that they must be incompetent, because not firmly based on what exists.  For, after all, we have no clue to why our common ideas based on existence do work.

Yet perhaps our ordinary ideas and the thought experiments are, in the end, similar and founded in the same way upon that fay creature, typicalness.  Then we should find the thought experiments to be less strange and our ordinary ideas to be more strange than we ever imagined.

The thought experiments are an algorithm for finding practical laws out of principal laws, by serendipity rather than by logic.  The workhorses will sometimes make use of wave functions just as the arabs do, and the distinction between practical and principal laws may not always be clear.  It is true then that this scheme is only loosely defined---it hangs after all on our powers of empirical and mathematical thinking.  We can never fully know what these powers are capable of.  Nonetheless the scheme does have design, enough to make it promising.

Such a system could have been spotted as a possible way to understand quantum mechanics around the time John von~Neumann presented his measurement theory---seventy years ago.  There has never been a hint of it.  It is far from what physicists think physics is or ought to be.

\numberedsectionz{4}{The probabilities of physics}

\noindent Probability has been a conundrum ever since it entered physics in the nineteenth century through the study of gases.  Quantum mechanics allows the computation of probabilities.  If the events limned by quantum mechanics are not real ones, whatever can its Born probabilities mean?  There is no chance in a quantum algorithm.  It is calculation.  One must draw conclusions from the algorithmic scenes, but what gamble is there in that?

A quantum mechanical calculation from a wave function will give you a set of possible observation results, each accompanied by its Born probability.  In your thinking to infer practical laws, of course you must take more seriously those observation results with the higher probability.  The Born probabilities act as weights.  But this is not yet satisfactory.  Weight is too vague.  How seriously should you take an observation result with a weight of one fiftieth?

We need some real probabilities here.  Here is how to get them.  Fancy that the whole set of possible observation results is not provided to you.  Instead, someone chooses just a few for you to see.  The choice is made completely at random.  The probability of choosing each outcome is to be found by using its Born probability made into a real probability by, say, rolling dice.  But you get only the several outcomes, without being told their odds.

See if you can figure out what laws you would {\it most likely\/} infer if this were done.  If you can do that, you are giving Born probabilities their right weight.

It is natural to use an {\it assortment\/} of wave functions to study some kind of event.  Not just one will be typical of it.  To choose the assortment, you are thrown on your judgment.  This will dilute probability, but make it more life-like.  For events fill our lives, but their outcomes rarely happen with a well-defined probability.

With a few kinds of event, then, even an assortment of wave functions, if the wave functions are at all appropriate, will define fairly definite probabilities for the several possible outcomes.  With the rest of events, an assortment will be less coherent in its implications, showing us probabilities only in the more nebulous sense of that word.

\numberedsectionz{5}{Measurement theory}

\noindent The measurement theory that von~Neumann blocks out in his {\it Mathematische Grundlagen der Quantenmechanik\/} of 1932 (von~Neumann [1955]\bracketsepz) lays bare the mathematical backbone of observation algorithms.  The theory is beautiful.  Nevertheless, two things von~Neumann did wrong.

Sure, and times of measurement do not belong in his theory.  They seem gauche.  By the precision of their definition they violate both the spirit of quantum mechanics and the practice of scientists.  And they are not modestly a time at a place, they are a hypersurface bisecting space-time.  If you consider measuring a property of anything larger than a particle, these hypersurfaces will portray a measurement of it that is nothing like a measurement any physicist has ever made.  The hypersurfaces are unreal.

They are also unnecessary.  Born probability is defined without using times of measurement.  Both the wave function and the measurement result can be evolved by the Hamiltonian, giving us two mathematical objects extending over all time.  The Born probability between them, defined in the usual way at one time, is the same defined at any other time---so in fact it is timeless.

If these objects tendered over all time, the wave functions and measurement results, are the natural basis of the theory---and its mathematics suggests this strongly---then, despite all custom, this theory must not be a theory of measurement.  It must be something else.  These objects have to do with no particular time.  But measurements, in general, will refer to a measured property as possessed by a physical system at some moment.

There are, to be sure, some measurements that are not tied to a moment.  To find constants of motion is an example.  Von~Neumann\rq s theory without times of measurement could describe that type of quantum measurement.  Yet these are only measurements by halves.  They have only some of the properties a good measurement should have.

Historically, Born found his probabilities in scattering theory.  He made no use of a time of measurement there.  This concept is a later intrusion.

It appears in the Schr\"odinger and Heisenberg representations.  Each gives you just one thing defined over all time, the wave function or the measurement result.  The other is given only at a certain instant.  (Actually, in the Heisenberg representation the evolution of the operator that stands for the variable to be measured is presented. But the operator defines the possible measurement results as eigenfunctions.)  Here would be a wonderful project for a historian, to discover why this was done.  The idea of measurement of position may have been central.

Quantum jumps, made into the \lq{}projection postulate\rq{}, are von~Neumann\rq s other lapse.  Jumps are without doubt everything to the Bohr atom; they played a vital role in the birth of quantum mechanics.  Too, the maiden success of Heisenberg\rq s new quantum theory was to define values for the \lq transition probabilities\rq---the strengths of Bohr\rq{}s spectral lines.

But such saltations do not fit into the foundations of von~Neumann\rq s theory.  Quantum jumps must be like the gushings of geysers, then.  They can be inferred from the theory, surely, but are not to be bolted right to its base.  

Von~Neumann left a record (\bracketsepz[1955]\bracketpagesepz pp.\abbrevfiguresepz \hbox{212--14}) of the origin in his mind of the projection postulate.  He tried to discover the significance of the celebrated Compton-Simon experiment of 1925 and came to see it so:

Photons of known energy, coming from a known direction, strike electrons at rest in a target.  The incoming momenta of the two colliding particles are had, then.  The energy and direction of flight of both particles after the collision are detected, so the outgoing momenta are found too.  Any three of these momenta settle the fourth---if conservation of momentum holds.

On this assumption, either of the measurements of the momentum of an outgoing particle determines, in conjunction with the two known incoming momenta, all aspects of the collision.  In particular, the momentum transfer between the two particles is fixed by this means (he considers the central line, which is the direction of this transfer).  The two determinations of the momentum transfer did in fact agree, since this is just another way of saying that momentum is conserved, which is what Compton and Simon found to be true.

The Compton-Simon experiment had famously shown the Bohr-Kramers-Slater theory of 1924 to be wrong.  By that theory momentum would be conserved on average, but not in each individual collision.  It was the quantum mechanics of Heisenberg that conserved momentum precisely.

Von~Neumann thought, though, that he had found an even deeper significance in the Compton-Simon experiment than the conservation of momentum.  For he saw each detection of an outgoing particle momentum as a {\it measurement of the transfer of momentum\/}.  One of the two measurements is made a little after the other, by \lq\lq{}usually about $10^{-9}$ to $10^{-10}$ seconds\rq\rq{} \hskip -.1em(\bracketsepz[1955]\bracketpagesepz p.\abbrevfiguresepz 213).  So the lesson of the Compton-Simon experiment is that in the brave new world of atoms a quick repeat of a measurement will get you the same number again.

He assessed on this basis the three rivals of the time, classical mechanics, the Bohr-Kramers-Slater theory, still quite fresh in mind, and Heisenberg{\rq}s promising new quantum mechanics.

According to classical mechanics, the result of a measurement is determined by the state of the system one is measuring.  The state will not change quickly, so if one measurement comes promptly after another, it must find the same value.  But experiments showed that classical mechanics is wrong.  And they suggested that the results of measurements were not determined, were a matter of chance.  The two new theories presumed this.

If measurement results can be a matter of chance, it might or might not be so that a closely following measurement will always arrive at the same outcome.  Von~Neumann\rq{}s analysis of the Compton-Simon experiment tied this to the conservation of momentum.  By the Bohr-Kramers-Slater theory, then, the following measurement need not agree.  By quantum mechanics, however, the second measurement must give the same result as the first---despite that the result of the first came only at random.  This latter way in measurement, he thought, is what nature has chosen---and with it quantum mechanics.

Then the course that has proven to be nature\rq{}s is more like classical mechanics than the Bohr-Kramers-Slater theory envisioned.  (But von~Neumann did not make this point, though others did.)  Momentum is precisely conserved.  And a measurement done near after another will have the same issue.  These will be defining qualities of the new quantum physics.

For this reason von~Neumann introduced his projection postulate (\bracketsepz[1955]\bracketpagesepz pp.\abbrevfiguresepz \hbox{214--23}, \hbox{347--58}, \hbox{417--21}).  A system quickly jumps into the eigenstate that corresponds to the value the measurement found for the variable.  (By the accepted theory, a measured value will be an eigenvalue of the operator representing the variable.)   When more than one state corresponds to this value, he said that the system will jump to one of them, but he could not say which \hbox{(pp.\abbrevfiguresepz 218, 348--49)}.  In this manner an immediate remeasurement of the variable must come out the same way, in agreement with his analysis of the Compton-Simon experiment (p.\abbrevfiguresepz 335).

But von~Neumann has been dreadfully careless.  The measurement theory he developed plumb invalidates his analysis of the Compton-Simon experiment.  His {\it analysis\/} says that the momentum transfer between the particles is measured.  This is something whose value can be known only if you know at least one of the two incoming momenta.  (In fact, he assumes knowledge of both incoming momenta.)  In quantum mechanical terms, in order to find the momentum transfer you require some knowledge of the incoming wave function.  But by his {\it theory\/}, you need not know anything about that wave function to know what the measurement result is.  Certainly momentum transfer has meaning, but it is not a measurable quantum variable!

Moreover, what is more seriously wrong than its origin, von Neumann{\rq}s projection postulate is utterly incompatible with the pith of his measurement theory.  Be a system given that interacts at various times with instruments, whose dials are seen.  His basic theory will insist on telling you what observations of the system this effects and what the probabilities of the results are.  There is just no room for further postulates in the matter.

It is like this.  The system and the instruments are combined into one wave function.  The observations will be of light coming from the dials. This light is part of the wave function, too.  The observations are also made into one.  That is to say, results of the light observations are deemed one combined result.  Formally, there is a single observation of a single wave function.

The way instruments touch the system may bespeak that by this means the {\it system\/} is observed in several ways, at different times.  But the algorithm is simple; as it sees it, it makes precisely one observation (of light given  off).  The algorithm allows no changes, lets no quantum jumps be put in.

The light observations do need to be, in a technical sense, compatible with each other.  Only then can they be made together and combined.  It is easy to find lots of compatible sightings of light. There is no problem here.

Von~Neumann would say that these observations of the light scattered from the dials are made simultaneously.  This need not mean that they are made at the same time.  There is no need to assign times to them at all.  Only the information they gather concerns us, not when the information is acquired.  Would it please, dream the light is intercepted light-years away.

But if no projection acts, a wave function will evolve into a linear combination of live-cat-with-unbroken-flask-in-a-box and dead-cat-with-broken-flask-in-a-box.  This need not be a calamity in a thought experiment, however.

At the start the wave function should clearly typify a live cat and undecayed atom.  As time passes the sense of the wave function will become less transparent to us, hinting at divergent possibilities.  The information held in the wave function at the start is less and less able to determine the situation.  Correspondingly, the congeries of possible sights produced by observation of the wave function becomes more diverse.  But each individual sight shows, within the reach of its view, a particular case, such as a live cat or a dead one.

As often said, for cats in boxes entanglement probably has no practical significance.  It will act as if a mixture.  But \lq practical significance\rq{} is happier in a thought experiment than in a definitive description of what exists!

So, were you to study Schr\"odinger\rq{}s cat by thought experiments, you will find, I am sure, that the idea that an unseen cat is either dead or alive works.  Though not quite perfectly.  Does a cat die just when its heart stops?

Musing that the cat is entangled will be possible too, but idle.  The dead-or-alive idea works better for us, and it is permitted.  Yet there may be special macroscopic circumstances that only a counterintuitive linear combination can twig.

It is mordant.  A young von~Neumann is deeply impressed by a profound discovery about measurement, found in his analysis of the Compton-Simon experiment.  He will not let go of it (\bracketsepz [1955]\bracketpagesepz pp.\abbrevfiguresepz 351--58, 417--21).  He mangles his measurement theory to fit.  There is no clearer, sadder example in the annals of science of deciding beforehand and being stubborn.  He killed his child.

Pauli did not help.  He put measurements into two classes, those that conformed with von~Neumann\rq s prescription and those that did not.  Maybe this was logical, but von~Neumann\rq s quickly repeated measurement idea was wrongheaded and fruitless, and it needed to be said.

Now it is time to tell about the good things in von~Neumann\rq s theory.  As a start, a little essay I wrote will be dropped in place here.  It will give a relaxed and spacious look at the thought experiments.

\numberedsectionz{6}{Typicalness, radiation, and inference}

\noindent Measurement is not the only way to garner physical information, photography is another.  But photography is physics\rq{} orphan.  A measurement gives direct access to physical reality; a photograph means nothing until it is interpreted, and that interpretation may involve subtleties of judgment that can never be thoroughly understood.

So physics is deemed to be grounded on measurement---but its topic may be photography.  Quantum mechanics has a theory of photography that is natural to the mathematics; its theory of measurement is still a worriment.  Quantum photography would be quite difficult to work with, its calculations tough.  On the other hand, the present theory of measurement, though easy to use, appears to be an oversimplification.

\numberedsubsectionz{6.1}{Quantum measurement}

\noindent We will look at measurement first.  Classical physicists proposed that any physical system always had accurately defined properties.  These were quietly thought to be, at least in principle, measurable at any time, so that nature might be an open book.

In the quantum Schr\"odinger representation, a system is defined by a time-dependent wave function.  Call it an {\it orbit\/}.  For any particular time, the orbit yields the state of the system.  But the state does not determine the value a physical variable has, as in classical mechanics. If you measure, you might get any of various results.

For physicists quickly forged an elementary theory of quantum measurement.  An operator represents a variable to be measured.  The operator\rq{}s eigenvalues are the values of the variable that may be found.  The eigenvector corresponding to a value gives, depending on the state of the system at the time, the probability of obtaining that value in a measurement.

\eject

The pioneers of quantum mechanics had found a simple analogy.  The soul of quantum mechanics will be that measurements of physical variables are to quantum mechanics what physical variables are to classical mechanics.  And they gave the probabilities of the results of hypothetical measurements at any precise time as sharp a definition as ever the variables of classical mechanics had.  The two theories were to be twins, if not in their meaning then in what seemed even more important, the exactness of their delineation of what they refer to.

But the potential measurements the pioneers posited are highly ideal, to the point of being fantastic.  In our world instruments make rough measurements, and they will not work at all except on a limited range of systems.  So this original theory of measurement, although some of the ideas in it may be important, is certainly not a definitive theory of measurement.

Yet if one improves the theory, making the measurements it outlines more realistic, they become sloppier, not at all like the precise values at precise times that variables have in classical mechanics.  When measurement theory gets better the pioneers\rq{} parallel must break down glaringly.

They gave the idea of measurement too much honor.  Too soon people did not feel free to innocently explore and find what is there in quantum mechanics\rq{} garden.

That serves as prologue to this story.

Von~Neumann (\bracketsepz[1955]\bracketpagesepz chap.\abbrevfiguresepz 6) elaborated the theory of measurement of~the pioneers by showing how a measuring instrument could be treated as a quantum system interacting with the system to be measured.  Since all measurements~are made by using instruments, measurement theory should then be what follows from this development of the original theory.  But there are changes to the original theory.  They are not modifications, they are revolutionary.

The first change:  In von~Neumann\rq{}s theory a measurement result is an eigenvector, not an eigenvalue.  By his analysis, what one can tell when an instrument is used and its dial observed, is just the probability that various dial readings will come about.  Vectors define these probabilities.  Eigenvalues have nothing to do with it.

Vectors can be developed by the Hamiltonian into orbits.  Then von~Neumann\rq{}s theory will have an orbit both for the system to be measured and for the observation result.  There is a resemblance to scattering theory.  

No times of measurement have to be specified, because Born probabilities are defined between two orbits without need to single out a time (or, it does not matter what time you use).  This is a relief, for the sharply defined times of the original theory never suited the sense of quantum mechanics.

The second change:  Vectors, or even density matrixes, as a rule are not able to express the information gathered with the help of an instrument.  A more general mathematical form we may call {\it co-orbits\/} (they will be time-dependent) is needed.  Born probabilities are now defined between an orbit and a co-orbit.  Co-orbits are like orbits, but can contain less information.  In the extreme case, no information is gathered by an observation at all.

The result is a theory of getting information from observations without making measurements.  For, typically, to observe by using an instrument cannot even nearly be characterized as a measurement of a variable of the system observed.

Also upset is the fundamental principle of classical physics, that physics is all about the states of physical systems.  Co-orbits are not orbits.  The twain can not both be states.

And we have only orbits and co-orbits, with no measurement reference times.  But this is not a tack new to physics.  Born\rq{}s 1926 paper on scattering (Born [1983]\bracketsepz) that introduced probability to the new quantum mechanics, contains no times of measurement.

\numberedsubsectionz{6.2}{Quantum photography}  

\noindent Look at how the camera works.  If a picture of a thing is to be taken, the distance of the camera from this source of light is not critical.  A camera at a greater distance will capture essentially the same picture as would a closer camera whose shutter was snapped slightly sooner.  Taken alone, when the shutter is snapped is then not acutely important either.

A camera may \lq{}measure\rq{} that the position of a photon is at a certain grain in the camera\rq{}s film at the instant the shutter is snapped.  But this event is not significant in itself.
 
To interpret the image in the photograph usefully, one must suppose the photon originated at a distance from the lens and traveled freely through clear space from the source.  From the position of the developed grain one can then get the direction the photon came from.  Its orbit!  The time of the snapping of the shutter and the distance of the camera together determine the phase of the photon (the timing of its motion, or how far ahead of or behind other photons traveling the same path this photon is), which is also an aspect of its orbit.  The wavelength (color) of the photon might be found by the camera too, and this too is part of the orbit.

Quantum mechanics is smoothly able to represent the kind of information a camera gathers in, though not by photon orbits.  Photon co-orbits are just the thing.  Remember, co-orbits are similar to orbits, except they can contain less information.

A camera photographs radiation.  The theory of the camera will then be somewhat like scattering theory, for scattered particles radiate away from their collision.  However, where scattering theory uses orbits both in and out, photography theory uses orbits in and co-orbits out.

Rather naturally, scattering theory has always been regarded as the study of the transition of particles from the state they had before the collision to the state they have afterward.  Photography theory cannot be about such transitions.  The spaces of orbits and co-orbits are not alike.

By von~Neumann\rq{}s theory subsystems with independent origins do not interact with each other in the far past, nor are they entangled with each other then.  Rather similarly, co-orbits belonging to diverse, but compatible, observations do not interact nor are entangled in the far future.  They represent elements of radiation that can be separately intercepted.

Radiation yields a bounty of co-orbits that do not interact in the future.  There can be a myriad of quantum observations of radiation that are all compatible with each other.  Formally, such a myriad will be one combined observation.

Quantum photography is not the theory of the snapshot, it is cinematography.  Observation of radiation is not inherently static.  In real life one can take a movie easily .  Its frames are very compatible observations.

The movie-like nature of quantum photography comes about like this.  A two-photon orbit may be composed from two one-photon orbits.  Say that each represents a photon traveling along a line, the two passing through a certain point about a second apart.  By a similar construction a two-photon co-orbit can be formed from two one-photon co-orbits.  The co-orbit can represent the information obtained by a camera there, if it shot two frames a second apart and each frame detected one photon.

But this co-orbit would need to show no information about photons between the two frames or outside the field of view of the camera, where the orbit shows an absence of photons instead.  This illustrates the necessity of the greater mathematical generality of co-orbits.

Quantum photography will now provide us with pictorial histories of things happening.  It is natural that it should.

\numberedsubsectionz{6.3}{Imaginary experience}

\noindent Do not seek the actual movies that quantum photography describes.  Those actual movies that would answer must be as rare and aloof as unicorns.  Still, quantum photography might tell us something sound about nature anyway.  For we are interested in seltzer water.

Observation of isolated systems tells us all we need to know about physics.  So we will make up a (real) freshly carbonated, isolated glass of seltzer.  It might be enclosed in a glass sphere along with a lamp.  We watch it.  We see bubbles forming, rising, and popping until the seltzer goes flat.  We have learned that seltzer fizzes.  Can quantum photography match this?

We should be able to turn up a quantum orbit that typifies, especially at some moment we choose, a glass of seltzer enclosed in a glass sphere with a lamp.  The glass sphere gives the seltzer an appropriate local environment (air pressure, but we have overlooked gravity).  The whole forms an isolated system, and this is what orbits seem made to describe.

Wave functions spread and entangle.  The orbit can represent a glass of seltzer nicely only for a certain period.  Then it will get muddy and muddier.  So our chosen time is when the orbit represents a glass of seltzer best.

There should be no light radiating away from the glass sphere at this special time.  Let the orbit produce its own radiation, as it will, lickety split.  Light {\it we\/} might put in {\it then\/} is our guess at what light the seltzer had emitted before the special time---a guess at what the seltzer looked like in the past.  In essence, we will have conjectured how the seltzer had behaved.  We ought not.

We need to find a suitable set of co-orbits too, one that is typical of the set of {\it all\/} possible movies that some camera might take at some given position and time with a certain number of frames, speed of the film, etc.  This set will be vast.  Almost all of these movies will be just visual noise.  Among the comparatively few that are not will be every kind of informative movie you can imagine, most all of them not about seltzer.

Now we combine this set of movie co-orbits with the seltzer orbit.  Each co-orbit will be given a probability by the orbit.  Those with the greatest probability (which will still be tiny for each) will show a glass of seltzer fizzing.  Quantum mechanics has just told us that seltzer fizzes.

\numberedsubsectionz{6.4}{Time reversal}

\noindent Fizzing is irreversible, while the orbit evolves reversibly.  Yet the two go together, because there is that special time.  The orbit is, in important ways, simple and unentangled then.  It will contain no inherently informative radiation.  That kind of radiation is generated by Hamiltonian evolution, and it will be entangled with the subsystem it emerged from.  There can be unentangled ambient radiation at the special time, but, just so that we may tell the story readily, we will start the orbit off with no radiation.

{\it After\/} any time when there is no light, the orbit\rq{}s core (the glass sphere with the seltzer) will radiate freely into space.  And there will be no light in space that might converge on the core later.  But what happens {\it before\/} that time?

Look at the orbit time-reversed.  The time of no light will have none still.  So the orbit going back will behave in the same general way.  Now reverse the orbit to forward again.  In this way we see that, at all times {\it before\/} the time of no light, the forward orbit contains light converging on the seltzer (and finely entangled with it!).  Nor can there be any radiating away, {\it before\/}.

The co-orbits will represent detection of light escaping from the seltzer, but not detection of light converging on it.  This is how a movie camera aimed at seltzer behaves.  Therefore the movies will be dark until that time when the orbit has clearest definition, the special time, then show fizzing.

An orbit typical of a glass of seltzer, if time-reversed, will become an orbit that is typical of a perhaps slightly different glass, as it happens.  If typical slight currents are rendered in the original, the reversed currents may not be typical currents, but it will be near enough like a typical seltzer.  So the reversed orbit will fizz in its own way.  (For clarity of reference, say that the reversed orbit fitzes.)  Of course, an orbit typical of some other kind of system might change behavior dramatically when time-reversed.

The undetected converging light in the first half of the original orbit will become the detected radiating light in the last half of the reversed orbit.  This light will make movie co-orbits depicting fitzing probable.

A quantum movie is not a movie that the reversed orbit produces and then the movie run backward, like the way classical mechanics would have it.  For each movie is patterned by just half of an orbit, the half with outgoing radiation.

The two half orbits handle a pair of cases that have precisely reversed movement just as they start.  A pair of classical motions that start thus can be welded back to back into one continuous motion that has no beginning and must be reversible.  So too with the half orbits, but a quantum pair of movies cannot be stitched together, and they will be depictions of irreversible happenings.

\numberedsubsectionz{6.5}{Defining the behavior that is typical of nature}

\noindent The full calculations of quantum photography are hopelessly hard.  Finding shortcuts and approximations will be very difficult.  It is not clear what will be possible for us.  But something will.

If the calculations were easy we would learn marvelously about nature\rq{}s behavior from our thought experiments.  Yet they are incurably just thought experiments.  Real systems do not have a time when they are best defined; figmental systems designed for quantum calculations must.

\eject

Thought experiments will not tell us plainly what to expect.  We are left to infer, however we can, a variety of theories from the movies (fizzing of seltzer, increase of entropy) that we may apply to our affairs, in the usual way with practical ideas.  These useful theories need not be classical.  Some can make use of quantum concepts, even wave functions.

We will make inferences from quantum sights as we do from the sights of our genuine experience, with the twist that we have the help of our knowledge of the mathematics that produces the movies.  This is a strange new science.  But it is probably enough like our customary empirical habits of thought that we can handle it.

Typicalness, radiation, and inference are the three timbers of imaginary experience.  For one, an orbit will not represent what exists.  Even though an orbit might contain all possible detail, we will choose an orbit to be typical of a kind of thing.  Orbits are explicitly good for naught else.  Next, we learn of typical behavior through the informing capacity of radiation.  Finally, the content of the orbits and the sights painted by the co-orbits are expressly a~riddle.  We must draw on all our cleverness to get out meaning from them.  None of this were we expecting to find in the rock bottom laws of nature

Indeed, no one would have dreamt that nature\rq{}s typical local behavior might be defined this way.  It may be nature\rq{}s way, all the same; it need only be effective at defining her typical behavior.  Nature does not even have~to~permit us to fully ferret out why her way of defining her passel of orderliness works.  Note that ruefully but well.

Whatever our misgivings, quantum photography does nevertheless sketch what nature may do.  Let there be a more fundamental way in which quantum mechanics defines nature\rq{}s local behavior, if you will.  It is constrained to agree with whatever quantum photography may tell us.

But I do not see how quantum photography, without some advance, can give us cosmological results.  We do have cosmological results, and here it might be that quantum photography does not plumb quantum mechanics\rq{} depths (still, I believe it does).

[A comment on this point:  Lay out a flat, classical space.  In the center will sit a quantum system which emits radiation.  Now modify this, I do not know how, so that the quantum system sits in quantized space but still sends its radiation out into a classical-like space.  Then invert the construction.  A~quantum system in quantized space surrounds a patch of classical space into which it spews radiation, which is observed.

The universe as a whole might be studied by thought experiments of this apt form.  In the classical patch of space picture astronomers looking out at and discovering the quantized universe.  The place where they dwell is not treated like the rest of the universe, but this is not fatal.  The vignette is only to be a thought experiment, not a mirror of reality.]

\numberedsectionz{7}{Von Neumann\rq{}s theory grounded on observation of radiation}

\noindent I would like to show in more detail how von~Neumann{\rq}s theory supports the thought experiments.  The ideas that will be brought forward ought to be based on clear mathematics, and I do not have that.  These are guesses, but good guesses, I think.  They anticipate the morphology of the theory when it is properly developed. 

By his projection postulate von~Neumann tried to mimic an ideal of classical physics, the measurement that does not disturb.  The classical property that is measured is not changed.  A quickly following measurement will then get the same result.  He could echo only the latter trait, and that not gracefully.

But quantum mechanics actually has an ideal both more realistic and more perfect.  When we catch the radiation that is freely given off by a quantum system, sent out into space never to return, we learn something of that beaming system without disturbing it at all.  Naturally, though, we will have disturbed the radiation we catch.  This will not concern us the least bit.

We can con a thing, too, by putting instruments onto it.  These may disturb the object.  But the information they gather will be sent away in radiation.  We observe this radiation without disturbing the object and instruments.

In sum, a magnificent way to learn of all things without bothering them is to monitor their radiation.  Among these things are observations themselves, both disturbing and undisturbing ones.  These observations to be studied may be fashioned by installing their instruments, and the instruments might observe radiation, as we do too, or be more meddling.

For these reasons, the algorithms that express von~Neumann\rq{}s theory will firstly sense radiation, and all will hinge on these primary observations.

\numberedsubsectionz{7.1}{Basics}

\noindent A physical system may be made up of two or more systems that began apart.  To show this, in the distant past the orbit of the group is (at least very nearly) a tensor product of orbits.  Each alone would represent a component system when alone.  Later these orbits may be interact and entangle.

Not every set of orbits for individual systems can be combined in this way.  For example, there may be two orbits that if put together by tensor product at any one time whatever, would interact and entangle at once.  When two orbits can be combined, we will say they are {\it past distinct\/}.  Orbits~$p$~and~$q$  combined are represented by a commutative and associative product called~$p \times q$.

This is not itself a tensor product.  The orbit has the tensor product form only in early times.  This product is more physics than mathematics, not always defined, and when defined, a little rough-hewn.

An observation may be effected by separately observing several things.  Were a contributory observation made alone, the information come by would be represented by a certain co-orbit.  The several things observed must be separately accessible, so these several co-orbits will not be interacting in late times.  The result of the combined observation, then, is that co-orbit with the form, throughout the future, of the tensor product of those contributory co-orbits.  They need not remain unentangled toward the past.  If co-orbits $S$~and~$T$ are {\it future distinct\/}, they combine into the product~$S \ctimescz T$.

The probability that an observation of a system will have a certain result is determined by the orbit of the system and the co-orbit of the result alone.  That is to say, it is not influenced by the method of observation, nor by what other possible results the observation may have.  The Born probability between an orbit~$p$ and co-orbit~$S$ is~\vex{.27mm}{.27mm}$\leftbornhspacez\bp pS \Sb\hspacerightbornz\bornsepz$\kern-.2em.

The co-orbits will be gathered into sets named {\it bindles\/}.  A bindle represents all the possible results of an observation.  For any orbit~$p$, the sum of the Born probabilities~\vex{.27mm}{.27mm}$\leftbornhspacez\bp pS \Sb\hspacerightbornz\bornsepz$\kern-.2em, for all the co-orbits~$S$ in a bindle~$\{S_i\}$, will be one.  The co-orbits in a bindle need not be future distinct from each other.  They can even be identical.  (Two separate possible results of an observation, which is to say two distinct kinds of event, might yield the same information.  A co-orbit represents information.)

Two bindles will be {\it consonant\/} if they represent the possible results of two observations that may be done together {\it without interfering at all with each other\/}.  We will see what that means shortly.  Each co-orbit in one bindle will be future distinct from each co-orbit in the other.  It follows, of course, that two co-orbits, one from each of the consonant bindles, may be combined by the product~\kern -.2em \ctimescz\kern -.2em.

But\kern .2em \ctimescz \kern .2em can {\it only\/} be used to combine co-orbits of observations that do not interfere with each other.  At least, the observations must not interfere with each other in the circumstance that they come out with the particular results that are to be combined.  This is a milder condition than that they never interfere with each other.

It is easy enough to see why use of~\kern -.1em \ctimescz \kern .15em must be limited in this way.  When two observations are made of a system $p$, the Born probability for the results $S$~and~$T$ to happen together, if it can be expressed in the form \vex{.27mm}{.27mm}$\leftbornhspacez\bp p\kern .02em (S \ctimescz T\kern .08em)\rightbornz\bornsepz$\kern-.2em, depends only on $p$, $S$, and~$T$.  That is not enough information to tell the probability that $S$~and~$T$ will happen together if either observation of~$p$ interferes with the other when these results occur.  The interference could take many forms, after all.  But if they do {\it not\/} interfere with each other, there can be only one probability  for the joint occurrence of their results $S$~and~$T$, when observing~$p$.  The role of \vex{.27mm}{-.9mm}$\leftbornhspacez\bp p\kern .02em (S \ctimescz T\kern .08em)\rightbornz$ is to say what it is.

\numberedsubsectionz{7.2}{Primitive observations}

\noindent We will build a set of bindles, designed to represent some observations of the radiation that springs from a central place.  These are the {\it primitive bindles\/}; they found the algorithm.  A {\it primitive observation\/} consists of its primitive bindle alone.  The other observations, which all have their own bindles, are constructed from primitive bindles and instruments.

Just for sake of simplicity, we will choose the primitive bindles to be all consonant with one another.  A bindle cannot be consonant with itself, however.  In this theory, the same observation cannot be made twice.

The Born probabilities of the primitive observations will satisfy this law:

\equationvspacez\vskip -.36mm

\equationz{0)}{$\leftbornhspacez\bp pS \Sb\hspacerightbornz\borneqsepz =\hskip.2em \sum _j \leftbornhspacez\bp p\kern .02em(S\ctimescz T_j)\rightbornz$} \equationforallsepz for all orbits $p$ considered,

\equationvspacez\vskip -.36mm

\noindent where $S$~is a co-orbit of one primitive observation and $\{T_j\}$~is the bindle of another.  This says that when observing any orbit~$p$, the observation~$\{T_j\}$ does not disturb the observation~$\{S_i\}$ when the result of the latter is~$S$.

For \vex{.36mm}{.45mm}$\leftbornhspacez\bp pS \Sb\hspacerightbornz$ is taken to mean the probability---when no other observation is made---that the observation~\vex{-36mm}{.45mm}$\{S_i\}$ will bring forth the result~$S$.  By similar logic, \vex{-36mm}{.45mm}$\leftbornhspacez\bp p\kern .02em (S\ctimescz T\kern .08em)\rightbornz$~is the probability that the two observations $\{S_i\}$ and~$\{T_j\}$ will have the results \vex{-36mm}{.45mm}$S$~and~$T$ together, these two observations alone being made.  It follows that \vex{-36mm}{.45mm}$\sum _j \leftbornhspacez\bp p\kern .02em (S\ctimescz T_j)\rightbornz$ will be the probability of~$S$ when there are just the two observations and the result of~\vex{-36mm}{.45mm}$\{T_j\}$ is unknown.  Equation~(0) says that this is the same as the probability of~$S$ when there is only the one observation.

This is the {\it definitive criterion of innocence\/} that one observation does not interfere with another.  Primitive bindles always meet it.  Hence \vex{.36mm}{.27mm}$\leftbornhspacez\bp pS \Sb\hspacerightbornz$~comes to mean the probability of~$S$ {\it regardless\/} of whether other primitive observations of radiation are done too, if their results are unknown.

\numberedsubsectionz{7.3}{The two division operations}

\noindent Besides the two \lq multiplications\rq, $\times$ \kern -.25em~and~\kern -.1em $\ctimescz$\kern -.2em, there are two \lq divisions\rq. The first division tells how the information provided by the outcome of one observation changes the probabilities of the potential results of any other observations that are made along with it.  We mean, here, that several observations must be mutually noninterfering if you can say they are made together, else you only tried to make them together.

The other division represents information about a system, when got by~way of a primitive observation aided by an instrument. The instrument is a second system that comes into being away from the first, then interacts with it.

People have used this latter division to express the information that is in a measurement result.  To get a result, they apply a measuring instrument to the examinee and then observe the instrument\rq{}s dial.

The division is actually broader in scope.  For example, the observation that sees the dial might see more, it might catch direct sight of the examinee.  And it even might not see the instrument, which has no dial.  This says it best: the primitive observation will win information about the examinee, and the instrument has an influence on this.  The instrument need not be the sole agency for the investigation.

Neither of these divisions is new.  They are implemented mathematically by partial traces.  See Karl Kraus\rq{}s {\it States, Effects, and Operations\/} (Kraus~[1983]\bracketsepz).  But they will be treated here in an uncommon manner.

\vskiplinez

\noindent {\it The first division\/}.  We have two primitive (and so mutually noninterfering) observations of a system whose orbit is~$p$.  There is a certain probability that the second observation will come to a result whose co-orbit is~$T$ when the first has resulted in~$S$.  This conditional probability is the prior probability that both will happen, which is~\vex{.27mm}{.73mm}$\leftbornhspacez\bp p\kern .02em (S\ctimescz T\kern .08em)\rightbornz\bornsepz$\kern-.2em, divided by the prior probability that $S$~will happen, which is~\vex{-36mm}{.36mm}$\leftbornhspacez\bp pS \Sb\hspacerightbornz\bornsepz$\kern-.2em.

We will define a new orbit which will yield all these new probabilities that are conditioned on the occurrence of~$S$.  It will be the orbit~$q$ for which \vex{.36mm}{.36mm}$\leftbornhspacez qT\hspacerightbornz\borneqsepz=\borneqsepz\leftbornhspacez\bp p\kern .02em (S\ctimescz T\kern .08em)\rightbornz\kern .2em/\kern .2em\leftbornhspacez\bp pS \Sb\hspacerightbornz\bornsepz$\kern-.2em, to hold for all the co-orbits~$T$ in the bindles of primitive observations that are consonant with the primitive observation whose result is~$S$.  These will be all the primitive observations other than the one whose result is~$S$---given our choice of a simple set of primitives. 

However, probability theorems are neater when a probability distribution is normalized, not to one, but to the probability that the circumstance will happen that makes the distribution germane.  Then every probability given by the distribution will be the joint probability that this circumstance, and the event the probability is linked to within the distribution, both happen.

A probability distribution normalized to one shows, instead, conditional probabilities that events will happen.  The condition, of course, is that the state of affairs that the distribution is designed to deal with has come about.

The same is true for orbits.  They, too, skip more nimbly if the probability distributions they define are normalized to the probability that the event that makes an orbit the right one will happen.  It does take getting used to.  One is accustomed to thinking in terms of conditional probabilities, which frankly posit that a significant event has occurred and go on from there.

Our orbit~$q$ becomes apt when $S$~happens, for $q$~gives probabilities that are predicated on that event.  The probability of~$S$ is~\vex{.18mm}{.18mm}$\leftbornhspacez\bp pS \Sb\hspacerightbornz\bornsepz$\kern-.2em.  We want the probability distributions that bindles create to be normalized to this.  So we will multiply $q$~by~\vex{.18mm}{-.9mm}$\leftbornhspacez\bp pS \Sb\hspacerightbornz$ to give us~$r$---if an orbit~$q$ is multiplied by a constant~$a$ to form the orbit~$aq$, it means that \vex{-3.6mm}{.27mm}$\leftbornz(aq)T\hspacerightbornz\borneqsepz =\borneqsepz\kern .05em a\kern -.1em \leftbornhspacez qT\hspacerightbornz$ for all the relevant co-orbits~$T$.  Real numbers between zero and one, like~$a$ here, I call {\it bambinos\/}.

We will say that \vex{.27mm}{1mm}$\leftbornhspacez\bp pS \Sb\hspacerightbornz$~is the {\it norm\/} of this orbit~$r$.  In symbols, $\|r\|$.  So 

\noindent\hfill$\|r\|\kern.2em=\borneqsepz\leftbornhspacez\bp pS \Sb\hspacerightbornz\bornsepz$\kern-.2em,\hfill $r\kern.2em=\kern.2em\|r\|q$,\hfill $1\kern.2em=\kern.2em\sum _j\leftbornhspacez qT_j\hspacerightbornz\bornsepz$\kern-.2em,\hfill\strut

\noindent\vex{1.2mm}{1.2mm}\hfill$\leftbornhspacez qT\hspacerightbornz\borneqsepz=\borneqsepz\leftbornhspacez\bp p\kern .02em (S\ctimescz T\kern .08em)\rightbornz\kern .2em/\kern .2em \leftbornhspacez\bp pS \Sb\hspacerightbornz\bornsepz$\kern-.2em.\hfill\strut

Therefore, for any bindle~$\{T_j\}$ consonant with \vex{-3.6mm}{.73mm}$\{S_i\}$, \kern.2em$\|r\|\kern.2em =\kern.2em \sum _j\leftbornhspacez rT_j\hspacerightbornz\bornsepz$\kern-.2em.  For~$T$s in these bindles, \vex{-3.6mm}{.73mm}$\leftbornhspacez rT\hspacerightbornz\borneqsepz=\borneqsepz\leftbornhspacez\bp p\kern .02em (S\ctimescz T\kern .08em)\rightbornz\bornsepz$\kern-.2em.  The orbit~$r$~gives, with a~$T$, the joint probability that $S$~and~$T$ result when $p$~is observed.

An orbit now has, in its norm, an extra item of information, the probability of the event that makes the orbit pertinent.  Just by dividing a new style orbit (like~$r$) by its norm, an old style orbit (like~$q$) is recovered.  It provides probabilities based on the premise that this event has happened.

The division~$p\odivbycz S$ shall mean~$r$.  Then \vex{.45mm}{.73mm}$\leftbornz(\kern .08em p\odivbycz S\kern .08em)T\hspacerightbornz\borneqsepz=\borneqsepz\leftbornhspacez\bp p\kern .02em (S\ctimescz T\kern .08em)\rightbornz$ for all the~$T$s in bindles consonant with~$\{S_i\}$.  And~\vex{-3.6mm}{.45mm}\kern.2em$\|\kern .08em p\odivbycz S\kern .08em\|\kern.2em=\borneqsepz\leftbornhspacez\bp pS \Sb\hspacerightbornz\bornsepz$\kern-.2em.

Orbit norm with division~\kern -.25em$\odivbycz$ \hskip -.2em can express Born probabilities, as just shown.  Then we may use the rewritten equation \vex{.45mm}{.45mm}\kern.2em$\|\kern .08em p\odivbycz (S\ctimescz T\kern .08em)\|\kern.2em=\kern.2em\|(\kern .08em p\odivbycz S\kern .08em)\odivbycz T\kern .08em\|$, if we wish. Its sides were switched for easier comparison with later equations.

\vskiplinez

\vskip-.4mm

\noindent {\it The second division\/}.  If we want to learn about a physical system, it may help to look after it has interacted with another system.  One system may be represented by the orbit~$p$ when it is alone, the other by~$q$.  They should have arisen independently of each other, so the two together will be~\kern.2em$p\times q$.

A primitive observation is made of the whole.  The probability of getting the result~$S$ is~\vex{.27mm}{.27mm}$\leftbornz(\kern .08em p\times q)S\hspacerightbornz\bornsepz$\kern-.2em.  The co-orbit~$S$ represents information gained about system and instrument.  No prior knowledge is supposed of either.

When the {\it instrument is known\/} (its orbit in the far past is~$q$) and the primitive observation of the system and instrument has a result, we gain information about the {\it system\/}.  This information will be expressed by a co-orbit~$U$\kern-.1em.  It will form a Born probability with the system\rq{}s orbit.

Mark that we look for a co-orbit~$U$ \kern-.1em to express the information, rather than, say, for the value of a variable.  It is a peculiarity of quantum mechanics that the information contained in an observation result takes the form of the specification of the probability of getting the result, for any possible orbit of a system observed.  This is what co-orbits provide.

What is telling, a co-orbit thereby sets new probabilities, for the potential results of other observations, that are conditioned on the result it represents.  The product~\kern -.2em$\ctimescz$ \hskip -.2em is important to this end, since it helps to define the joint probabilities of results of mutually noninterfering observations.

The co-orbit~$U$ \kern-.1em we seek, and the primitive observation result~$S$, are two aspects of the same event and must have the same probability of coming up.  Thus~\vex{.27mm}{-.9mm}$\leftbornhspacez\bp pU \Sb\hspacerightbornz\borneqsepz=\borneqsepz\leftbornz(\kern .08em p\times q)S\hspacerightbornz\bornsepz$\kern-.2em.  Since $U$~\kern-.1em depends on $q$~and~$S$, let \kern.2em$S\cdivbyoz q$ \kern.2em mean~$U$\kern-.1em.  Then~$\leftbornhspacez\bp p\kern .02em (S\cdivbyoz q)\rightbornz\borneqsepz=\borneqsepz\leftbornz(\kern .08em p\times q)S\hspacerightbornz\bornsepz$\kern-.2em, for all the~$p$\kern .07em s that have a genesis unlinked with~$q$.  Here we may say, if we wish, that \vex{.55mm}{.36mm}\kern.2em$\|(\kern.08em p\times q)\odivbycz S\kern.08em\|\kern.2em=\kern.2em\|\kern.08em p\odivbycz(S\cdivbyoz q)\|$.

The whole bindle~$\{S_i\}$ of the primitive observation (which observes instrument and examinee) is changed, by the instrument whose orbit is known to be~$q$, into a new bindle~$\{S_i\cdivbyoz q\}$. This is for a different observation (it observes the examinee) that is not primitive.  It is an {\it instrumented observation\/}.  The {\it instrument\/},~$q$, and the {\it look\/},~$\{S_i\}$, fashion it.

The norm of~$q$ may not be one.  The definition of~\kern.2em$S\cdivbyoz q$ \kern.2em then gives us that $\leftbornhspacez\bp p\kern .02em (S\cdivbyoz q)\rightbornz\borneqsepz=\borneqsepz\leftbornz(\kern .08em p\times q)S\hspacerightbornz\borneqsepz=\kern.2em\| q\|\kern .1em \leftbornz(\kern .08em p\times \doubleunderlinez{q}\kern .08em)S\hspacerightbornz\bornsepz$\kern-.2em, \vex{.73mm}{.73mm}for orbits~$p$ independent of~$q$.  The symbol~\vex{-3.6mm}{.55mm}$\doubleunderlinez{q}$~means $q$~normalized to one; in other words, $q$~equals $\| q\|$ times~$\doubleunderlinez{q}$.  The law \kern.2em$p\times aq\kern.2em =\kern.2em a(\kern .08em p\times q)$ \kern.2em was used.  This~\kern.2em$S\cdivbyoz q$ \kern.2em defines joint probabilities.  They are for the happening of the event that makes the orbit~$q$ represent the instrument, together with the observation result that \kern.2em$S\cdivbyoz q$\kern.2em~represents.

We said, when introducing the concept of bindle, that if $\{S_i\}$~is a bindle then \vex{.45mm}{.55mm}\kern.2em$\sum_i \leftbornhspacez\bp pS_i \Sb\Sb\hspacerightbornz\borneqsepz=\kern.2em1$, \kern.2em for any orbit~$p$ (of norm one).  This no longer need be true, since co-orbits have been generalized.  For example, for the bindle~$\{S_i\cdivbyoz q\}$ we have \vex{.64mm}{1mm}\kern.2em$\sum _i \leftbornhspacez\bp p\kern .02em (S_i\cdivbyoz q)\rightbornz\borneqsepz=\kern.2em\sum _i \leftbornz(\kern .08em p\times q)S_i\hspacerightbornz\borneqsepz=\kern.2em\| q\|\kern .1em \sum _i \leftbornz(\kern .08em p\times \doubleunderlinez{q} \kern.08em)S_i\hspacerightbornz\borneqsepz\borneqsepz =\kern.2em\|q\|$.

So now the sum may be any bambino.  This will be the bindle~$\{S_i\}$\rq{}s {\it norm\/}.  In symbols, \vex{.31mm}{.6mm}\kern.2em$\|\{S_i\}\|\kern.2em=\kern.2em\sum_i\leftbornhspacez\bp pS_i \Sb\Sb\hspacerightbornz\bornsepz$ for any relevant~$p$ of norm one.

A co-orbit~$S$ should be seen as a part of the bindle it is in.  One can normalize the co-orbit by multiplying it by the reciprocal of the norm of its bindle.  (When a co-orbit is multiplied by a constant~$a$, then \vex{.55mm}{.55mm}$\leftbornhspacez\bp p\kern .02em (aS)\rightbornz\borneqsepz=\borneqsepz\kern .05em a\kern -.1em \leftbornhspacez\bp pS \Sb\hspacerightbornz\bornsepz$ for all relevant orbits~$p$.)  A normalized co-orbit,~\vex{-3.6mm}{.36mm}$\doubleunderlinez{S\vrule width 0pt depth .1ex}$\kern .07em, gives the right probability for a result when events have made its bindle the right one.  Naturally, the norm of a bindle of normalized co-orbits is one.

When we equated \vex{.55mm}{.73mm}\kern.2em$\sum _i \leftbornhspacez\bp p\kern .02em (S_i\cdivbyoz q)\rightbornz$ to~$\|q\|$ above, we assumed that the norms of $p$~and~$\{S_i\}$ are one.  One will often want to take some orbits and bindles for granted:  the conditions for their use hold, and their norms are one.  These orbits and bindles are {\it pegged\/}.

The orbits~\kern.2em$p\odivbycz S$ \kern.2em and co-orbits~\kern.2em$S\cdivbyoz p$ \kern.2em have been defined in terms of the Born probabilities they define.  They have been given no other form, and no Hamiltonian evolution.  That they need to have in a complete theory, in order that the two multiplications may be defined for them, and not least to hint to us what these orbits and co-orbits might mean as representatives, in some fashion, of physical things.

\vskip -3.8mm

\numberedsubsectionz{7.4}{Four equations}

\noindent The two divisions have been termed divisions in order to draw on analogies with arithmetic.  Here are four equations that look arithmetical.

\equationvspacez

\equationz{1)}{$p\odivbycz (S\ctimescz T\kern .08em)\kern.2em=\kern.2em(\kern .08em p\odivbycz S\kern .08em) \odivbycz T$}

\equationvspacez

Two primitive observations of an examinee have been made.  Their results are now to be taken into account.  This equation states that the new probabilities based on this information, for the possible results of a third primitive observation, can be calculated equally well in either of two ways.  You may combine the two observation results into one and so take them into account together, getting the orbit~\kern.2em$p\odivbycz (S\ctimescz T\kern .08em)$.  Or you may take them one at a time, getting first the orbit~\kern.2em$p\odivbycz S$ \kern.2em and from that the orbit~\kern.2em$(\kern .08em p\odivbycz S\kern .08em) \odivbycz T$.  You can then calculate the Born probability (given that $S$~and~$T$ happened) of a result of a third observation, by using its co-orbit and either of these two equal orbits normalized.

The norms of the two orbits in equation~(1) did not help to define the probability of a result of a third observation.  But the equation does say the norms are equal, and this has its own meaning.  The equal norms imply that there are two ways to calculate the probability that the results $S$~and~$T$ will happen together.  You may find the probability of the two combined.  Or you may figure the probability of one result and multiply that by the probability that the other will happen, given that the first did.  In formulas, \vex{.09mm}{.27mm}$\leftbornhspacez\bp p\kern .02em (S\ctimescz T\kern .08em)\rightbornz$ equals $\leftbornhspacez\bp pS \Sb\hspacerightbornz$ (this is~\kern.2em$\|\kern .08em p\odivbycz S\kern .08em\|$, \kern.2em remember) times~$\leftbornz(\kern .08em \doubleunderlinez{p\odivbycz S}\kern .08em)T \hspacerightbornz\bornsepz$\kern-.2em.

\equationvspacez

\equationz{2)}{$S\cdivbyoz(\kern .08em p\times q)\kern.2em=\kern.2em(S\cdivbyoz p)\cdivbyoz q$}

\equationvspacez

Two instruments can work with a primitive observation to make an instrumented one.  The instruments can be introduced en~bloc or one at a time.  In the first method, the two are simply combined into one instrument.  In the second method, we use the instrument~$p$ with the primitive bindle~\vex{0mm}{.18mm}$\{S_i\}$ to make an intermediary instrumented observation whose bindle is~$\{S_i\cdivbyoz p\}$.  The examinee of this observation is the second instrument,~$q$, and the system we are interested in looking at, taken together.  Then we can use~$q$ as the instrument, and the bindle of the intermediary instrumented observation as the look, to make the ultimate instrumented observation.

\numberedsubsubsectionz{7.4.1}{When instrumented observations do not interfere}

\noindent Equality of the norms of the orbits on either side of equation~(3) below says that the probability of a result of the look observation is the same as the probability of the corresponding result of the instrumented observation, or \vex{.18mm}{.18mm}$\leftbornz(\kern .08em p\times q)S\hspacerightbornz\borneqsepz=\borneqsepz \leftbornhspacez\bp p\kern .02em (S\cdivbyoz q)\rightbornz\bornsepz$\kern-.2em.  That is always true.  Not so for equality of the two orbits themselves.  This is, rather, a condition of especial interest.

\equationvspacez

\equationz{3)}{$(\kern .08em p\times q)\odivbycz S\kern.2em=\kern.2em p\odivbycz(S\cdivbyoz q)$}

\equationvspacez

\vskip-.9mm

Envisage a space~$x$ of orbits within which the orbit~$p$ of a system to be observed must lie.  There will be two observations of this examinee.  One is instrumented, composed of its instrument~$q$ and its look, a primitive bindle~$\{S_i\}$.  The instrument hatches apart from the whole examinee space.  And there is, too, a primitive observation, with bindle~$\{T_j\}$.

The primitive observation will never interfere with the instrumented one.  It will not foil that other primitive observation, the look of the instrumented observation, nor push instrument or examinee around.

But under what circumstances will the instrumented observation not interfere with the primitive one?  Suppose the instrumented observation is made, its look results in~$S$, and this is minded.  The probability of a result~$T$ of the primitive observation may then differ from what it would have been if the primitive observation had been made by itself.

It seems reasonable to think that the primitive observation has not been interfered with (when $T$~turns out to be its result) if this is so:  That change in the probability of~$T$ depended only on the information about the examinee that the instrumented observation supplied---namely, the observation\rq{}s result~\kern.2em$S\cdivbyoz q$.  (That is to say, the change did not depend on the means of the instrumented observation---its instrument~$q$ or look~$\{S_i\}$.)  Call this yardstick the {\it criterion of the import of the observation result by itself\/} for noninterference by an instrumented observation.

If the criterion of import is not satisfied, then the presence of the instrument must have affected the chances of the primitive observation\rq{}s result.  The primitive observation might have seen the instrument.  Or it could have seen something in the examinee that was changed by the examinee\rq{}s interaction with the instrument.

The criterion of import is satisfied, when the instrumented observation results in~\kern.2em$S\cdivbyoz q$ \kern.2em and the primitive observation in~$T$, if the observations comply with the following condition:

\equationvspacez

\noindent

\equationz{3a)}$\leftbornz(\kern .09em p\kern .02em\times\kern .02em q)(S\ctimescz\kern .0em T\kern .13em)\kern .02em\rightbornz\borneqsepz=\borneqsepz  \leftbornhspacez\leftbracketz(\kern .08em p\times q)\odivbycz S\hspacerightbracketz\varfrombrackz T\hspacerightbornz\borneqsepz=\borneqsepz \leftbornhspacez\leftbrackethspacez p\odivbycz(S\cdivbyoz q)\rightbracketz\varfrombrackz T\hspacerightbornz\borneqsepz=$

\twixtequationpartsvspacez

\equationz{}$\leftbornhspacez\kern .08em p\varfrombrackz\leftbracketz(S\kern -.02em\cdivbyoz\kern -.02em q)\kern -.02em\ctimescz\kern -.02em T\kern .02em \hspacerightbracketz\kern -.03em \hspacerightbornz$ \equationforallsepz for all $p$ in~$x$.

\equationvspacez

Be sure to spot that the orbit~\vex{-3.6mm}{.18mm}\kern.2em$p\odivbycz(S\cdivbyoz q)$ \kern.2em when paired with the co-orbit~$T$ to form a Born probability, yields the joint probability of $S$~and~$T$, not the conditional probability of~$T$~given~$S$.  To get the conditional probability one must divide by the norm of~\kern.2em$p\odivbycz(S\cdivbyoz q)$.  Its norm is \vex{.27mm}{.27mm}\kern.2em$\| \kern .08em p\odivbycz(S\cdivbyoz q)\|\kern.2em=\borneqsepz\leftbornhspacez\bp p\kern .02em (S\cdivbyoz q)\rightbornz\borneqsepz=\borneqsepz\leftbornz (\kern .08em p\times q)S\hspacerightbornz\bornsepz$\kern-.2em, \vex{-3.6mm}{.36mm}the probability of~$S$.

The extreme right-hand side of equation~(3a) can rightly use~\kern-.1em$\ctimescz$ \hskip-.1em because here the instrumented observation\rq{}s result~\kern.2em$S\cdivbyoz q$ \hskip .2em satisfies the criterion of import with respect to the primitive observation\rq{}s result~$T$.  So $p$,~$T$, and~\kern.2em$S\cdivbyoz q$ do supply all the information necessary to determine the Born probability.

There is a second plausible idea for when an instrumented observation does not interfere with a primitive one.  It is that the mere presence of the instrument will not affect the probability of a result of the primitive observation. The result of the instrumented observation is ignored here.

The {\it criterion of transparency of the instrument to the other observation\/} is its name.  This expresses it, when the primitive observation results in~$T$:

\equationvspacez

\equationz{3b)}{$\leftbornz(\kern .08em p\times q)T\hspacerightbornz\borneqsepz=\borneqsepz\leftbornhspacez\bp pT\hspacerightbornz$ \equationforallsepz for all~$p$ in~$x$ ($q$~is pegged).}

\equationvspacez

Here is a sketch of a proof that meeting the criterion of import for {\it every\/} result of an instrumented observation, plus meeting the criterion of innocence, implies satisfaction of the criterion of transparency.  This sequence of transformations shows that \vex{.27mm}{-.9mm}$\leftbornz(\kern .08em p\times q)T\hspacerightbornz$~equals~$\leftbornhspacez\bp pT\hspacerightbornz\bornsepz$\kern-.2em, for all~$p$ in~$x$:

\equationvspacez

\equationz{\letterz{a\kern .01em}}{\kern-.2em$\leftbornz(\kern .08em p\times q)T\hspacerightbornz$}   

\twixtequationsvspacez

\equationz{\letterz{b\kern -.04em}}{\backabitz$\sum _i\foreabitz\kern-.2em \leftbornz (\kern .08em p\times q)(S_i\ctimescz T\kern .08em)\kern -.005em \rightbornz$}

\twixtequationsvspacez

\equationz{\letterz{c\kern .005em}}{\backabitz$\sum _i\foreabitz\kern-.2em \leftbornhspacez\leftbracketz(\kern .08em p\times q)\kern.01em \odivbycz \kern.015em S_i\kern .02em \hspacerightbracketz\varfrombrackz \kern .02em T\kern .1em \hspacerightbornz$}

\twixtequationsvspacez

\equationz{\letterz{d\kern -.02em}}{\backabitz$\sum _i\foreabitz\kern-.2em \leftbornhspacez\leftbrackethspacez\kern .04em p\kern.01em \odivbycz \kern.01em (S_i\cdivbyoz q\kern.005em)\kern .04em \rightbracketz\varfrombrackz \kern .02em T\kern .1em \hspacerightbornz$}

\twixtequationsvspacez

\equationz{\letterz{e\kern .02em}}{\backabitz$\sum _i\foreabitz\kern-.2em \leftbornhspacez\bp p\varfrombrackz\leftbracketz\kern -.03em(S_i \cdivbyoz q)\kern -.02em\ctimescz T \hspacerightbracketz\kern -.07em \hspacerightbornz$}

\twixtequationsvspacez

\equationz{\letterz{f\kern .15em}}{\kern-.2em$\leftbornhspacez\bp pT\hspacerightbornz$}

\equationvspacez

We go from~(a) to~(\kern.04em b\kern-.01em) because the criterion of innocence holds between primitive bindles.  From~(\kern.04em b\kern-.01em) to~(e) when $p$~and~$q$, all the results~$S_i$ of the look, and the result~$T$ satisfy equation~(3a).  And from~(e) to~(\kern.04em f\kern.12em) when the instrumented observation satisfies the criterion of innocence with respect to the primitive result~$T$.  This step expects, too, that the norm of~$q$ is one.

\numberedsubsubsectionz{7.4.2}{When observations are subsidiary to instrumented ones}

\noindent We will retain, for the purpose of understanding equation~(4\kern.03em) below, the space~$x$ for the orbits of examinees, the instrumented observation with instrument~$q$ and look~$\{S_i\}$, and the primitive observation with bindle~$\{T_j\}$.

\equationvspacez

\equationz{4)}{$(S\ctimescz T\kern .08em)\cdivbyoz q\kern.2em=\kern.2em S\cdivbyoz (q\odivbycz T\kern .08em)$}

\equationvspacez

The result of a primitive observation is {\it subsidiary\/} to an instrumented observation if, when you do not know the result of the look of the instrumented observation, the primitive observation has provided no information about the space of orbits that the instrumented observation is observing.  A subsidiary observational result may nevertheless be helpful when you do know the result of the instrumented observation\rq{}s look.

Heeding the subsidiary observation may modify the probabilities of the results of the instrumented observation.  Since the subsidiary result says nothing about the examinee, it seems plausible that it must say something about the instrument.  Then the whole impact on the instrumented observation ought to be effected by a change of the orbit of its instrument.  When this can be done the {\it criterion of bearing through the instrument alone\/} is met---a touchstone for subsidiary observational results.

If the primitive observation {\it had\/} seen into the space~$x$, then surely its result would have held too much information for this maneuver to work.  The observation could peek into the space to be examined by seeing an examinee, or just by seeing the instrument too late, after the instrument had interacted with an examinee---or even merely had had a chance to interact with an examinee had one been there.

The criterion of bearing is satisfied, should the primitive observation result in~$T$ and the look of the instrumented observation in~$S$, when the following condition is in force: 

\vskip -.93mm

\equationvspacez

\noindent

\equationz{4a)}$\leftbornz\kern .04em(\kern .12em p\kern -.02em\times \kern -.02em q)\kern -.06em(S\kern -.06em \ctimescz\kern -.06em T\kern .08em)\kern -.02em \rightbornz\borneqsepz=\borneqsepz\leftbornhspacez\bp p\varfrombrackz\leftbracketz(S\kern -.05em \ctimescz\kern -.03em T\kern .08em)\cdivbyoz q\hspacerightbracketz \hspacerightbornz\borneqsepz=\borneqsepz \leftbornhspacez\bp p\varfrombrackz\leftbrackethspacez S\cdivbyoz (q\kern -.05em \odivbycz \kern -.03em T\kern .08em)
\rightbracketz\hspacerightbornz\borneqsepz=$

\twixtequationpartsvspacez

\equationz{}$\leftbornhspacez\leftbrackethspacez\kern -.05em p\kern -.02em\times\kern -.02em(\kern .02emq\kern -.03em \odivbycz \kern -.03em T\kern .09em)\rightbracketz\varfrombrackz S\kern .04em \hspacerightbornz$ \equationforallsepz for all~$p$ in~$x$.

\equationvspacez

\vskip -.93mm

The Born probability that \vex{-3.6mm}{.18mm}\kern.3em$S\cdivbyoz (q\kern -.05em \odivbycz \kern -.03em T\kern .08em)$\kern.3em~and~$p$ define is the joint probability for $S$~and~$T$ to happen.  It is not the conditional probability of~$S$~given~$T$.

Whether or not $\{T_j\}$~is subsidiary, taking the occurrence of~$T$ into consideration will modify the instrumented observation into something like a new observation.  The \lq{}bindle\rq{} of this new \lq{}observation\rq{} is~\vex{.27mm}{.45mm}\kern.2em$\{(S_i\kern -.06em \ctimescz \kern -.02em T\kern .08em)\cdivbyoz q\}$.

But, \vex{-3.6mm}{.55mm}\kern.2em$\sum _i \leftbornhspacez\bp p^\prime\varfrombrackz\leftbracketz(S_i\kern -.06em \ctimescz \kern -.02em T\kern .08em)\cdivbyoz q\hspacerightbracketz \hspacerightbornz\borneqsepz=\kern.2em\sum _i \leftbornz(\kern .08em p^\prime\kern -.18em \times q)(S_i\kern -.06em \ctimescz \kern -.02em T\kern .08em)\rightbornz\borneqsepz= \borneqsepz\leftbornz (\kern .08em p^\prime\kern -.18em \times q)T\hspacerightbornz\bornsepz$\kern-.2em, if $p^\prime$ is any orbit of norm one in the space~$x$.  Thus in the general case, the \lq{}norm\rq{} of this \lq{}bindle\rq{} depends on which orbit~$p^\prime$ is used to define it.  An observation, however, really should not presuppose that we know the examinee{\rq}s orbit.  It has to count on the orbit\rq{}s lying within a certain space~$x$, though.

If we do have a proper observation, then its norm will be~$\leftbornz (\kern .08em p\times q)T\hspacerightbornz\bornsepz$\kern-.2em, the probability of~$T$.  When~\vex{.18mm}{.27mm}$S\cdivbyoz (q\kern -.05em \odivbycz \kern -.03em T\kern .08em)$, which we are taking to be equal to \vex{-3.6mm}{.18mm}\kern.1em$(S\kern -.05em \ctimescz \kern -.03em T\kern .08em)\cdivbyoz q$ \kern.1em here, is normalized in this way, it and~$p$ together give the probability of~$S$ given~$T$.

The extreme right-hand side of equation~(4a) implies that use of the multiplication~$\times$ is justified.  This is to say that $p$~and~\kern.2em$q\odivbycz T$ \kern.2em were solitary in the past, just as $p$~and~$q$ were.  But this is a matter I have given little thought to.

Here is another idea for when a primitive observation is subsidiary to an instrumented one.  Only the instrument and whatever is in the space~$x$ are present, and the probability of the result of the primitive observation does not depend on what is in~$x$.  Then the {\it criterion of total invisibility of the space of examinees to the primitive observation\/} is fulfilled.  (The primitive observation cannot see into the space~$x$, not even with the help of the instrument.)  It is expressed thus, when the result of the primitive observation is~$T$:

\equationvspacez

\noindent
\equationz{4b)}{$\leftbornz(\kern .08em p\times q)T\hspacerightbornz\borneqsepz=\borneqsepz\leftbornhspacez qT\hspacerightbornz$ \equationforallsepz for all~$p$ in~$x$ (the~$p$\kern .07em s are pegged).}

\equationvspacez

This is a sketch of an argument, for each~$p$ in~$x$, that satisfaction of the criterion of bearing with respect to {\it every\/} possible result of the look implies satisfaction of the criterion of total invisibility:

\equationvspacez

\equationz{\primedletterz{\normaltype a$^\prime$}}{\kern-.2em$\leftbornz(\kern .08em p\times q)T\hspacerightbornz$}

\twixtequationsvspacez

\equationz{\primedletterz{\normaltype b$^\prime$}}{\backabitz$\sum_i \foreabitz\kern-.2em \leftbornz(\kern .1em p\kern .05em \times\kern .05em q\kern .025em)(S_i\kern .03em \ctimescz\kern .03em T\kern .08em)\rightbornz$}

\twixtequationsvspacez

\equationz{\primedletterz{\normaltype c$^\prime$}}{\backabitz$\sum_i \foreabitz\kern-.2em \leftbornhspacez\bp p\varfrombrackz\leftbracketz(S_i\ctimescz T\kern .08em)\cdivbyoz q\hspacerightbracketz \hspacerightbornz$}

\twixtequationsvspacez

\equationz{\primedletterz{\normaltype d$^\prime$}}{\backabitz$\sum_i \foreabitz\kern-.2em \leftbornhspacez\bp p\varfrombrackz\leftbrackethspacez\kern .06em S_i\kern .04em \cdivbyoz\kern .04em (\kern .06em q\odivbycz T\kern .08em)\kern .1em \rightbracketz \hspacerightbornz$}

\twixtequationsvspacez

\equationz{\primedletterz{\normaltype e$^\prime$}}{\backabitz$\sum_i \foreabitz\kern-.2em \leftbornhspacez\leftbrackethspacez\kern .045em p\times (\kern .065em q\odivbycz  T\kern .085em)\kern .1em \rightbracketz\varfrombrackz\kern .035em S_i\kern .035em\hspacerightbornz$}

\twixtequationsvspacez

\equationz{\primedletterz{\normaltype f\kern .08em$^\prime$}}{$\kern-.2em\leftbornhspacez qT\hspacerightbornz$}

\equationvspacez

We go from~(a$^\prime$\kern.05em) to~(\kern.05em b$^\prime$\kern.03em) because the criterion of innocence holds between primitive bindles.  From~(\kern.05em b$^\prime$\kern.03em) to~(\kern.02em e$^\prime$\kern.03em) when $p$~and~$q$, all the results~$S_i$ of the look, and the result~$T$ satisfy equation~(4a).  And from~(\kern.02em e$^\prime$\kern.03em) to~(\kern.05em f\kern.08em$^\prime$\kern.03em) because the norm of~$p$ is one and the norm of~\kern.2em$q\kern .03em \odivbycz \kern .03em T$ \kern.2em is~$\leftbornhspacez qT\hspacerightbornz\bornsepz$\kern-.2em.\vex{.18mm}{.55mm}

\indent\vex{-3.6mm}{.82mm}\hbox{\hbox to 35.5mm {Equation (3a), \kern.05em if \kern.05em we \kern.05em wish, \kern.05em is}\kern 8.7mm
\hbox{$\kern.05em\|(\kern .09em p\times q)\odivbycz(S\ctimescz T\kern .08em)\|\kern.15em=\kern.15em\|\kern .08em p\odivbycz\leftbracketz(S\cdivbyoz q)\ctimescz T\hspacerightbracketz\kern .1em \|$.}}

\indent\vex{-3.6mm}{.55mm}\hbox{\hbox to 35.5mm {Equation (4a) can be put so:}\kern 8.7mm
\hbox{$\kern.05em\|(\kern .08em p\times q)\odivbycz(S\ctimescz T\kern .08em)\|\kern.15em=\kern.15em\|\kern .11em \leftbrackethspacez\kern .06em p\times(\kern .06em q\odivbycz T\kern .13em)\kern .02em \rightbracketz\odivbycz S\kern .12em\|$.}}

In this survey the scope of the definitions and equations has been in several ways unnecessarily limited to primitive observations.  This was done out of conservatism.  Primitive observations are simpler than instrumented ones.  I~stuck my toe in just far enough to test the water.

There are important structural departures from von~Neumann\rq{}s theory in the theory outlined here.  Here orbits and co-orbits have restricted domains.  (Those co-orbits for which an orbit may offer a Born probability are its domain.  The domains of co-orbits are analogous.)  For example, orbits of the form~$p\odivbycz S$ may act only with co-orbits from the bindles of observations that are mutually noninterfering with the observation that produced the result~$S$.  There will be reasons for even narrower domains.

Another departure:  The light that radiates away from the source in a certain solid angle (and there may be no light there) is a typical subsystem that a primitive observation might observe.  Or, for a better example, take the incoming light in the field of view of a camera.  This is quite different from a subsystem in von~Neumann\rq{}s sense.  Identical particles need to be handled, too.  A new kind of tensor product is needed for the orbits and co-orbits.  Actually, the observations should be implemented with quantum field theory.

\numberedsectionz{8}{Conclusion}

\noindent There is no way to know what treasures may lie in the thought experiments until they are explored.  But they do hold promise to unveil the precarious affiliation between the make-up of the orbits and the make-up of reality.  This is what I mean:

Orbits that might typify a snippet of existence, are the slightest part of the ocean of chaotic nonsense that the set of all orbits represents.  There is a problem of fishing for those efficacious orbits, then.  The division~\kern -.2em $\odivbycz$\kern -.2em~could help here by a process of annealing.  For example, we might choose an orbit that is not very \lq{}real\rq{} because the objects it represents are not much localized.  After some observations are made, an orbit expressing the results of these observations may represent the objects as better localized.  This would be imperfectly done, of course, because only what can be seen would be bettered.  But it would be a hint at what a \lq{}real\rq{} orbit should be like.  And orbits might be improved in other ways by observation.  Certainly, a linear combination of a dog and a cat would be resolved.  Maybe one could show that thermodynamic states should be Gibbsian ensembles.

\vskip 1cm

\leftskip=6.3mm
\rightskip=6.3mm
\noindent
There is something ineffable about the real, something occasionally described as mysterious and awe-inspiring; the property alluded to is no doubt its ultimacy, its spontaneity, its failure to present itself as the perfect and articulate consequence of rational thought. On the other hand mathematics, and especially geometry, have exactly those attributes of internal order, the elements of predictability, which reality seems to lack. How do these incongruous counterparts of our experience get together?

\vskip -1.4mm

\hfill Henry Margenau (\bracketsepz[1949]\bracketsepz)

\vfill
\eject

\leftskip=0mm
\rightskip=0mm

\addendumz

\noindent The thought experiments can in principle solve the problem classical mechanics could not, namely to show that belief in themselves is useful.

For you can do the following:  Make up a wave function typical of an astronomical gas cloud.  Then find movies made probable by the wave function that show a solar system forming.  Look for the development of life and of intelligent creatures on a planet.  By observation learn their language and so their thoughts.  See if they have conceived of quantum thought experiments and whether this seems to help them flourish.

Of course you cannot really do this.  And any sort of try at a shortcut would meet formidable difficulties.

\vskip -3mm

\acknowledgmentsz

\noindent The inspiration for the ideas in this paper principally was John von~Neumann\rq{}s book, {\it Mathematical Foundations of Quantum Mechanics\/} (\bracketsepz[1955]\bracketsepz).  The clear review that  Karl Kraus\rq{}s book, {\it States, Effects, and Operations\/} (\bracketsepz[1983]\bracketsepz), offers of the essentials of the matter and their mathematical underpinnings, gave me more confidence in what I was thinking.  My appreciation of the problem of contrary-to-fact conditionals is entirely due to N.~David Mermin (\bracketsepz[1990]\bracketsepz), who worked with Henry Stapp{\rq}s idea (\bracketsepz[1971]\bracketsepz), whom John Stewart Bell inspired.

\vskip -3mm

\referencesz

\referenceitemz{Born, M. [1983]: \lq{}On the quantum mechanics of collisions\rq{}, translated by Wheeler and Zurek, in J.~Wheeler and W.~Zurek ({\it eds\/}), {\it Quantum Theory and Measurement\/}, Princeton, NJ: Princeton University Press, \hskip .2em pp.~\hbox{52--55}.}
\referenceitemz{Kraus, K. [1983]: {\it States, Effects, and Operations\/}, Berlin: Springer-Verlag.}
\referenceitemz{Margenau, H. [1949]: \lq{}Einstein\rq{}s conception of reality\rq{}, in P.~A. Schilpp ({\it ed\/}), {\it Albert Einstein: Philosopher-Scientist\/}, La Salle, IL: Open Court,\kern3em \break \kern .2em Vol.\kern -.1em~1, \hskip .2em p.~250.}
\referenceitemz{Mermin, N.~D. [1990]: \lq{}Can you help your team tonight by watching on TV? More experimental metaphysics from Einstein, Podolsky, and Rosen\rq{}, in {\it Boojums All the Way Through\/}, Cambridge: Cambridge University Press,\hskip .2em pp.~\hbox{95--109}.}
\referenceitemz{Stapp, H. P. [1971]: \lq{}S-Matrix Interpretation of Quantum Theory\rq{}, Physical Review D3, \hskip .2em pp.~\hbox{1303--1320}.}
\referenceitemz{von~Neumann, J. [1955]: {\it Mathematical Foundations of Quantum Mechanics\/}, translated by R.~T. Beyer, Princeton, NJ: Princeton University Press.}
\referenceitemz{}

\vskip 2mm

donaldamccartor@earthlink.net

\end{document}